\documentclass[twocolumn,twocolappendix]{aastex631}

\shorttitle{JWST Atmospheric Reconnaissance of TRAPPIST-1~${\rm b}$}
\shortauthors{Lim et al.}

\usepackage{multirow}
\usepackage{tablefootnote}
\usepackage{soul}
\usepackage{amsmath}

\begin{document}

\title{Atmospheric Reconnaissance of TRAPPIST-1~b with JWST/NIRISS: Evidence for Strong Stellar Contamination in the Transmission Spectra
}

\author[0000-0003-4676-0622]{Olivia Lim}
\affiliation{Trottier Institute for Research on Exoplanets and Department of Physics, Universit\'{e} de Montr\'{e}al, Montreal, QC, Canada}

\author[0000-0001-5578-1498]{Bj\"{o}rn Benneke}
\affiliation{Trottier Institute for Research on Exoplanets and Department of Physics, Universit\'{e} de Montr\'{e}al, Montreal, QC, Canada}

\author[0000-0001-5485-4675]{Ren\'{e} Doyon}
\affiliation{Trottier Institute for Research on Exoplanets and Department of Physics, Universit\'{e} de Montr\'{e}al, Montreal, QC, Canada}

\author[0000-0003-4816-3469]{Ryan J. MacDonald}
\altaffiliation{NHFP Sagan Fellow}
\affiliation{Department of Astronomy, University of Michigan, 1085 S. University Ave., Ann Arbor, MI 48109, USA}

\author[0000-0002-2875-917X]{Caroline Piaulet}
\affiliation{Trottier Institute for Research on Exoplanets and Department of Physics, Universit\'{e} de Montr\'{e}al, Montreal, QC, Canada}

\author[0000-0003-3506-5667]{Étienne Artigau}
\affiliation{Trottier Institute for Research on Exoplanets and Department of Physics, Universit\'{e} de Montr\'{e}al, Montreal, QC, Canada}

\author[0000-0002-2195-735X]{Louis-Philippe Coulombe}
\affiliation{Trottier Institute for Research on Exoplanets and Department of Physics, Universit\'{e} de Montr\'{e}al, Montreal, QC, Canada}

\author[0000-0002-3328-1203]{Michael Radica}
\affiliation{Trottier Institute for Research on Exoplanets and Department of Physics, Universit\'{e} de Montr\'{e}al, Montreal, QC, Canada}

\author[0009-0005-6135-6769]{Alexandrine L'Heureux}
\affiliation{Trottier Institute for Research on Exoplanets and Department of Physics, Universit\'{e} de Montr\'{e}al, Montreal, QC, Canada}

\author[0000-0003-0475-9375]{Lo\"{i}c Albert}
\affiliation{Trottier Institute for Research on Exoplanets and Department of Physics, Universit\'{e} de Montr\'{e}al, Montreal, QC, Canada}

\author[0000-0002-3627-1676]{Benjamin V. Rackham}
\altaffiliation{51 Pegasi b Fellow}
\affiliation{Department of Earth, Atmospheric and Planetary Sciences, Massachusetts Institute of Technology, Cambridge, MA 02139, USA}
\affiliation{Kavli Institute for Astrophysics and Space Research, Massachusetts Institute of Technology, Cambridge, MA 02139, USA}

\author[0000-0003-2415-2191]{Julien de Wit}
\affiliation{Department of Earth, Atmospheric and Planetary Sciences, Massachusetts Institute of Technology, Cambridge, MA 02139, USA}

\author[0000-0001-6758-7924]{Salma Salhi}
\affiliation{Trottier Institute for Research on Exoplanets and Department of Physics, Universit\'{e} de Montr\'{e}al, Montreal, QC, Canada}
\affiliation{Department of Physics and Astronomy, University of Calgary, 2500 University Drive NW, Calgary, Alberta T2N 1N4, Canada}

\author[0000-0001-6809-3520]{Pierre-Alexis Roy}
\affiliation{Trottier Institute for Research on Exoplanets and Department of Physics, Universit\'{e} de Montr\'{e}al, Montreal, QC, Canada}

\author[0000-0001-6362-0571]{Laura Flagg}
\affiliation{Astronomy Department, Cornell University, Ithaca, NY 14853, USA}
\affiliation{Carl Sagan Institute, Cornell University, Ithaca, NY, USA}

\author[0000-0002-5428-0453]{Marylou Fournier-Tondreau}
\affiliation{Trottier Institute for Research on Exoplanets and Department of Physics, Universit\'{e} de Montr\'{e}al, Montreal, QC, Canada}

\author[0000-0003-4844-9838]{Jake Taylor}
\affiliation{Department of Physics, University of Oxford, Parks Rd, Oxford OX1 3PU, UK}
\affiliation{Trottier Institute for Research on Exoplanets and Department of Physics, Universit\'{e} de Montr\'{e}al, Montreal, QC, Canada}

\author[0000-0003-4166-4121]{Neil J. Cook}
\affiliation{Trottier Institute for Research on Exoplanets and Department of Physics, Universit\'{e} de Montr\'{e}al, Montreal, QC, Canada}

\author[0000-0002-6780-4252]{David Lafreni\`{e}re}
\affiliation{Trottier Institute for Research on Exoplanets and Department of Physics, Universit\'{e} de Montr\'{e}al, Montreal, QC, Canada}

\author[0000-0001-6129-5699]{Nicolas B. Cowan}
\affiliation{Department of Physics, McGill University, Montr\'{e}al, QC, Canada}
\affiliation{Department of Earth and Planetary Sciences, McGill University, Montr\'{e}al, QC, Canada}

\author[0000-0002-0436-1802]{Lisa Kaltenegger}
\affiliation{Carl Sagan Institute, Cornell University, Ithaca, NY, USA}
\affiliation{Cornell Center for Astrophysics and Planetary Science, Cornell University, Ithaca, NY 14853, USA}
\affiliation{Astronomy Department, Cornell University, Ithaca, NY 14853, USA}

\author[0000-0002-5904-1865]{Jason F. Rowe}
\affiliation{Department of Physics and Astronomy, Bishop's University, 2600 Rue College, Sherbrooke, QC J1M 1Z7, Canada}

\author[0000-0001-9513-1449]{N\'{e}stor Espinoza}
\affiliation{Space Telescope Science Institute, 3700 San Martin Drive, Baltimore, MD 21218, USA}
\affiliation{Department of Physics \& Astronomy, Johns Hopkins University, Baltimore, MD 21218, USA}

\author[0000-0003-4987-6591]{Lisa Dang}
\affiliation{Trottier Institute for Research on Exoplanets and Department of Physics, Universit\'{e} de Montr\'{e}al, Montreal, QC, Canada}

\author[0000-0002-7786-0661]{Antoine Darveau-Bernier}
\affiliation{Trottier Institute for Research on Exoplanets and Department of Physics, Universit\'{e} de Montr\'{e}al, Montreal, QC, Canada}




\begin{abstract}

TRAPPIST-1 is a nearby system of seven Earth-sized, temperate, rocky exoplanets transiting a Jupiter-sized M8.5V star, ideally suited for in-depth atmospheric studies. Each TRAPPIST-1 planet has been observed in transmission both from space and from the ground, confidently rejecting cloud-free, hydrogen-rich atmospheres. Secondary eclipse observations of TRAPPIST-1~b with JWST/MIRI are consistent with little to no atmosphere given the lack of heat redistribution. Here we present the first transmission spectra of TRAPPIST-1~b obtained with JWST/NIRISS over two visits. The two transmission spectra show moderate to strong evidence of contamination from unocculted stellar heterogeneities, which dominates the signal in both visits. The transmission spectrum of the first visit is consistent with unocculted starspots and the second visit exhibits signatures of unocculted faculae. Fitting the stellar contamination and planetary atmosphere either sequentially or simultaneously, we confirm the absence of cloud-free hydrogen-rich atmospheres, but cannot assess the presence of secondary atmospheres. 
We find that the uncertainties associated with the lack of stellar model fidelity are one order of magnitude above the observation precision of 89\,ppm (combining the two visits). Without affecting the conclusion regarding the atmosphere of TRAPPIST-1~b, this highlights an important caveat for future explorations, which calls for additional observations to characterize stellar heterogeneities empirically and/or theoretical works to improve model fidelity for such cool stars. This need is all the more justified as stellar contamination can affect the search for atmospheres around the outer, cooler TRAPPIST-1 planets for which transmission spectroscopy is currently the most efficient technique.

\end{abstract}



\section{Introduction} \label{sec:intro}

The TRAPPIST-1 system is host to seven transiting planets with masses and radii indicating a mostly rocky composition with, possibly, volatiles such as a surface water layer \citep{gillon_temperate_2016,gillon_seven_2017,luger_seven-planet_2017,agol_refining_2021}. The relatively small host star is a Jupiter-sized M8-type dwarf \citep{liebert_ri_2006,agol_refining_2021}, boosting the amplitude of potential spectral signatures from planetary atmospheres seen in transmission. The TRAPPIST-1 planets orbit their star in a compact configuration, allowing more transit observations within a given amount of time relative to planets with longer orbital periods. All these features combined make TRAPPIST-1 ideally suited to study multiple Earth-sized, rocky exoplanets from the same environment with some in the habitable zone of the system \citep{gillon_trappist-1_2020}. 

However, as an ultracool dwarf, TRAPPIST-1 likely stayed in the pre-main sequence phase for hundreds of millions of years. During the pre-main sequence phase, the TRAPPIST-1 planets were subjected to energetic radiation from the star, enhancing atmospheric escape processes \citep[e.g.,][]{wordsworth_water_2013,wordsworth_abiotic_2014,luger_extreme_2015,bolmont_water_2017,bourrier_temporal_2017,roettenbacher_stellar_2017,turbet_review_2020} accentuated by the proximity of the planets to the host. Even to this day, TRAPPIST-1 still emits in X-ray at a luminosity similar to that of the Sun despite its smaller bolometric luminosity \citep{wheatley_strong_2017}. The total EUV energy received by the TRAPPIST-1 planets over their lifetime ranges from 10 to 1000 times that received by Earth, depending on the planet \citep{fleming_xuv_2020,birky_improved_2021}. Flares were also observed on TRAPPIST-1 with $K2$ at frequencies of approximately 0.02--0.5 per day, each event releasing $10^{30}$--$10^{33}$\,erg in energy \citep{vida_frequent_2017}.

A first step towards studying the potential atmospheres of the TRAPPIST-1 planets is to assess their presence. Each TRAPPIST-1 planet has been observed in transmission by space- and ground-based observatories and these studies confidently rejected cloud-free, hydrogen-rich atmospheres \citep{de_wit_combined_2016,de_wit_atmospheric_2018,wakeford_disentangling_2019,gressier_near-infrared_2021,garcia_hstwfc3_2022,moran_limits_2018}. Secondary eclipse observations of TRAPPIST-1~b with the Mid-InfraRed Instrument \citep[MIRI,][]{bouchet_mid-infrared_2015} aboard JWST enabled the measurement of the dayside brightness temperature of the planet, $T_{\rm B} = 508^{+26}_{-27}\,$K, consistent with little to no heat redistribution to the nightside and a near-zero Bond albedo \citep{greene_thermal_2023}, ruling out CO$_2$-dominated atmospheres at surface pressures of 10 and 92\,bar, along with O$_2$-dominated atmospheres with 0.5\,bar of CO$_2$ at surface pressures of 10 and 100\,bar. A complementary analysis of these emission data \citep{ih_constraining_2023} rejected atmospheres with at least 100\,ppm CO$_2$ with surface pressures above 0.3\,bar, as well as a Mars-like, pure CO$_2$ atmosphere with a surface pressure of 6.5\,mbar. Secondary eclipse observations of TRAPPIST-1~c with JWST/MIRI also disfavor thick, CO$_2$-rich atmospheres \citep{zieba_no_2023}, but certain scenarios involving less CO$_2$ remain consistent with the data \citep{lincowski_potential_2023}. While secondary eclipse observations are efficient at assessing the presence of an atmosphere on the inner, hotter TRAPPIST-1 planets, applying this technique to the outer, cooler planets of the system is observationally expensive, even with JWST \citep{lustig-yaeger_detectability_2019}. Transmission spectroscopy is more sensitive to thinner atmospheres as it probes the atmosphere through the longer slant transit geometry, and is thus currently the only applicable technique to probe the potential atmospheres of the cooler TRAPPIST-1 planets, including those in the habitable zone \citep{gillon_trappist-1_2020}.

Here we present the first transmission spectra of TRAPPIST-1~b from JWST using the Near-Infrared Imager and Slitless Spectrograph \citep[NIRISS,][]{doyon_near_2023} in the Single-Object Slitless Spectroscopy (SOSS) mode \citep{albert_near_2023}. A challenge affecting transmission spectroscopy of planets orbiting active stars is stellar contamination \citep[e.g.,][]{sing_hubble_2011,mccullough_water_2014,rackham_access_2017,rackham_transit_2018,rackham_sag21_2023}. When a planet transits a star with surface heterogeneities (spots or faculae), the flux from the entire visible stellar disk, which is typically taken as the reference light source, may not be representative of the flux that actually gets occulted by the planet during its transit, on the transit chord. When not accounted for, this transit light source (TLS) effect contaminates the transmission spectrum with features from the \textit{stellar} spectrum, potentially biasing in the planetary atmosphere retrieval. The NIRISS transmission spectra of TRAPPIST-1~b show strong evidence of stellar contamination, but nonetheless allow us to reject certain atmospheric scenarios.

\section{Observations} \label{sec:observations}

The goals of the JWST TRAPPIST-1 atmospheric reconnaissance program 
(JWST GO-2589, PI Lim) are to assess the presence of planetary atmospheres in the TRAPPIST-1 system, to investigate the impact of unocculted stellar heterogeneities on JWST transmission spectra, and to compare the performance of NIRISS \citep{doyon_near_2023} and NIRSpec \citep{jakobsen_near-infrared_2022}, in the context of detecting and characterizing the atmospheres of exoplanets orbiting late M dwarfs. As part of this reconnaissance program, we observed two transits of TRAPPIST-1~b on 2022 July 18 and 2022 July 20 with NIRISS in SOSS mode \citep{albert_near_2023}, covering the 0.6 -- 2.8\,$\mu$m wavelength domain at a resolving power of approximately $700$. Each visit spans from $2.59$\,hours before ingress to $1.25$\,hours after egress, for a total of $4.44$\,hours per visit. The observation windows were selected to minimize contamination from field stars given the slitless nature of the instrument. Each observation was performed in a single exposure of 153 integrations of 1.65\,minutes (18 groups per integration, duty cycle of 89.5\%) using the SUBSTRIP256 subarray and the NISRAPID readout pattern. We scheduled the two transits based on predicted mid-transit times BJD$_{\rm TDB}=$~2,459,779.210512\,days (2022 July 18) and BJD$_{\rm TDB}=$~2,459,780.721350\,days (2022 July 20), and transit duration $T_{\rm dur} = 36.06\,$minutes \citep{agol_refining_2021}.

\section{Data Reduction} \label{sec:floats}

We reduced the data from both visits using three separate pipelines, \texttt{supreme-SPOON} \citep{feinstein_early_2022,coulombe_broadband_2023,radica_awesome_2023}, \texttt{NAMELESS} \citep{feinstein_early_2022,coulombe_broadband_2023}, and \texttt{SOSSISSE}, 
to ensure redundancy and verify the consistency of our results. The \texttt{supreme-SPOON} and \texttt{NAMELESS} pipelines have already been presented in the literature, so we include only a short summary of each in Appendix~\ref{app:reduction}, noting steps that differ from previous works. Since \texttt{SOSSISSE} is a new method, we present it in full in Appendix~\ref{app:reduction}. Unless otherwise stated, quantitative results and figures presented hereafter are shown for the \texttt{NAMELESS} reduction given its more conservative error bars, but we note that they are similar for the other methods (see Appendix~\ref{subapp:reduction_consistency}). 

\begin{figure*}
    \centering
    \includegraphics[height=.76\textheight]{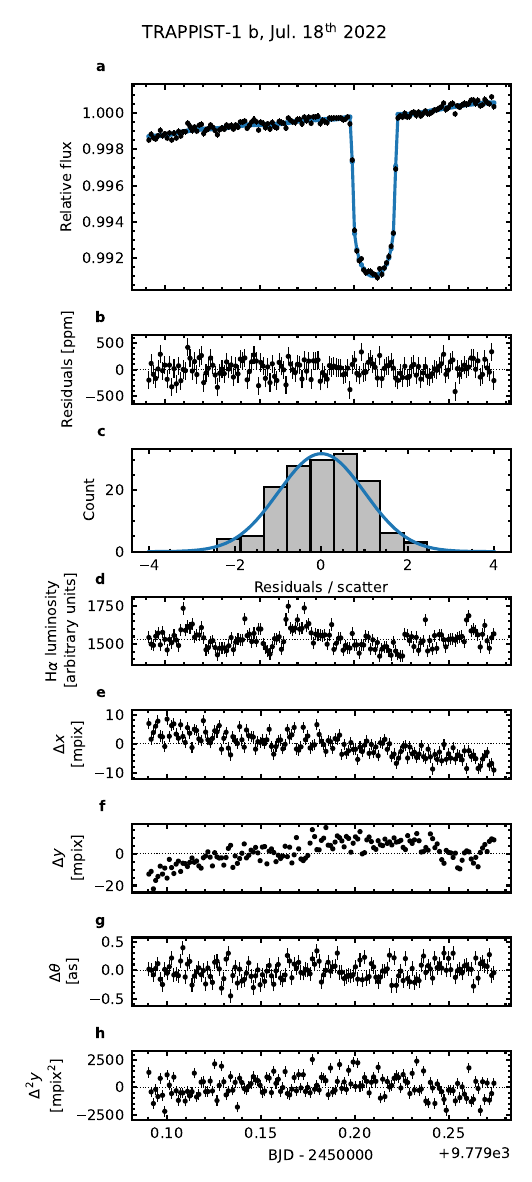}
    \includegraphics[height=.76\textheight]{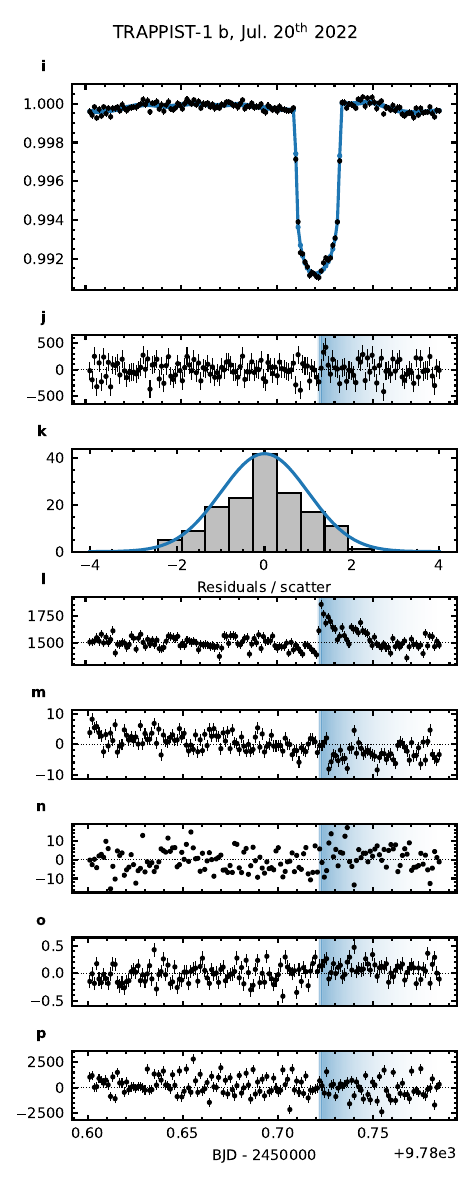}
    \caption{NIRISS/SOSS broadband light curve fits, H$\alpha$ integrated flux, and spectral trace morphology metrics. Panels (a)--(h) correspond to visit 1 and panels (i)--(p), to visit 2. All error bars are the 1-$\sigma$ uncertainties. (a) \& (i) Observed light curve (black points) and best-fit model (blue curve). (b) \& (j) Light curve residuals from the best-fit model. (c) \& (k) Histogram of the light curve residuals normalized by the scatter (bars) and a Gaussian distribution (blue curve) for comparison. (d) \& (l) Time series of the integrated flux in the H$\alpha$ stellar line. (e) \& (m) Amplitude of the trace displacement in the spectral direction ($b(t)$ in text). (f) \& (n) Amplitude of the trace displacement in the cross-dispersion direction ($c(t)$). (g) \& (o) Amplitude of the trace rotation ($d(t)$). (h) \& (p) Amplitude of the second derivative of the trace profile in the cross-dispersion direction ($e(t)$). For visit 2, the shaded blue domain highlights the stellar flare. The jump in the broadband flux (j) and in the H$\alpha$ integrated flux (l) is not linked to any anomaly in the spectral trace morphology ((m)--(p)).}
    \label{fig:wlc_fit}
\end{figure*}

The extracted light curves from the two visits exhibit several indicators of stellar activity (Section~\ref{sec:lcfit}, Figures~\ref{fig:wlc_fit}(a) and \ref{fig:wlc_fit}(i), and Appendix~\ref{app:stelvar}). The most obvious indicator is a temporal slope in the light curve of the first visit and a curvature in the second visit, which we interpret as stellar flux modulation.

\section{Transit light curve fitting} \label{sec:lcfit}

With the extracted spectra, we performed a global analysis of the broadband and spectroscopic transit light curves with ExoTEP \citep[][see Appendix~\ref{app:lcfit}]{benneke_spitzer_2017,benneke_sub-neptune_2019,benneke_water_2019}. 

On top of long-timescale flux modulations such as a slope or a curvature, the light curves from both visits exhibit correlated noise on shorter timescales (e.g., around the transit egress of visit 2; see Figure~\ref{fig:wlc_fit}(i)). These temporal flux variations are correlated with the wavelength-integrated flux in the H$\alpha$ stellar line (see Figure~\ref{fig:wlc_fit}(l) and Appendix~\ref{app:stelvar}), and are not correlated with any spectral trace morphological metrics we computed (e.g., trace displacement; see Figures~\ref{fig:wlc_fit}(m)--\ref{fig:wlc_fit}(p)), which is suggestive of a \textit{stellar}, not \textit{instrumental}, nature of the correlated noise. We interpret the bump in the second transit as a stellar flare. Had we not detected the H$\alpha$ line variability, we could have mistaken this structure in the second transit for a spot-crossing event \citep[$3.5$-$\sigma$ detection using \texttt{SPOTROD}\footnote{\url{https://github.com/bencebeky/spotrod}},][]{beky_spotrod_2014}. The H$\alpha$ line is thus an important proxy for stellar flares, and its variability is more unambiguously detected when its spectral resolution is higher because the line is then less diluted, giving an edge to NIRISS/SOSS relative to NIRSpec/Prism (see also Howard et al., submitted).

\begin{figure*}
    \centering
    \includegraphics[width=\textwidth]{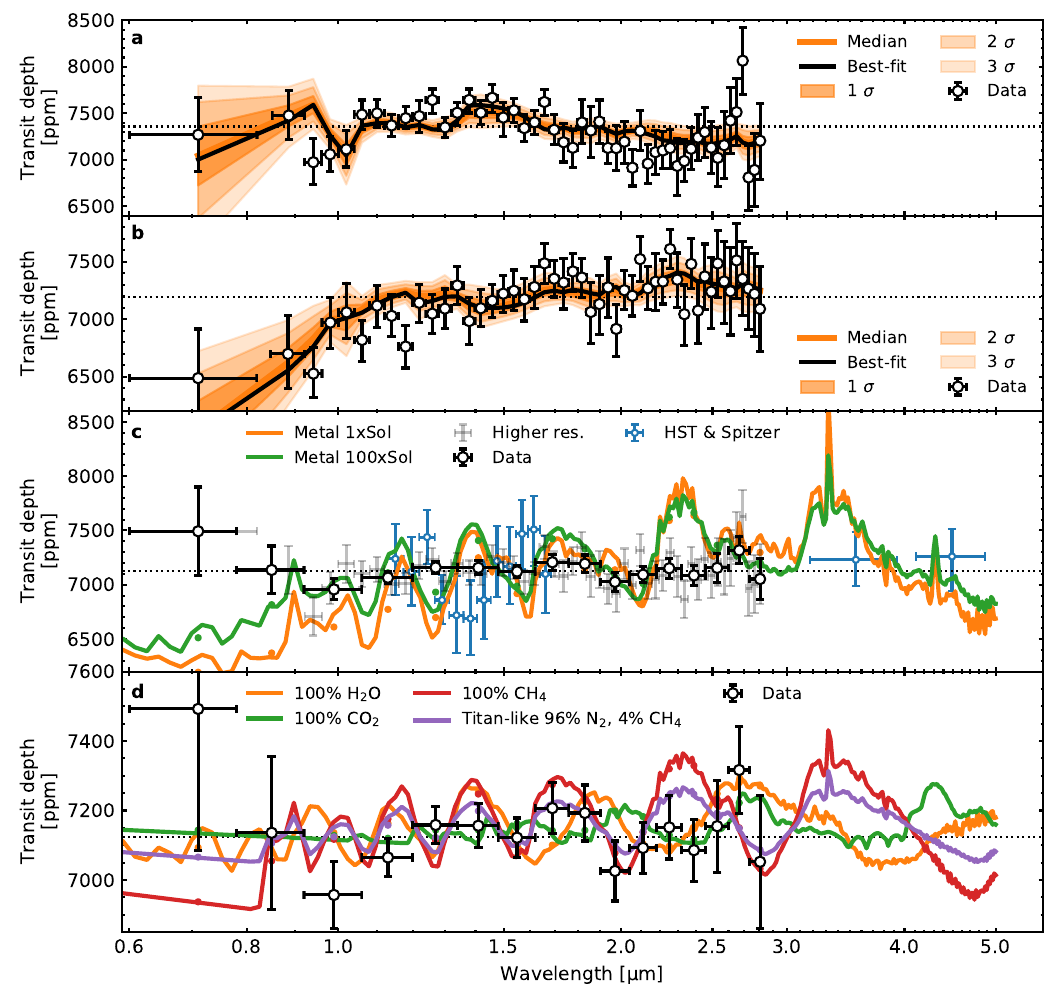}
    \caption{NIRISS/SOSS transit spectrum of TRAPPIST-1~b compared to stellar contamination and atmosphere models from the sequential analysis. Black circles are the SOSS transit spectra, either from visit 1 (a), visit 2 (b), or from both visits combined ((c)--(d)). In panels (c) and (d), the transit spectra are corrected from stellar contamination. Dashed lines are the error-weighted mean transit depths. Vertical error bars are the 1-$\sigma$ uncertainties. Horizontal error bars represent the extent of each spectral bin. (a)--(b) Comparison between the transmission spectrum of each visit to its best-fit/median stellar contamination model (black/orange curves) and uncertainties (shaded regions). (c) Grey, thin error bars are the SOSS transit depths at higher spectral resolution. Blue points are the HST/WFC3 and Spitzer/IRAC transit depths \citep{de_wit_combined_2016,de_wit_atmospheric_2018,zhang_near-infrared_2018,ducrot_trappist-1_2020}, vertically shifted to match the median of the SOSS data. Clear hydrogen-rich models in orange and green can be ruled out at $29$ and $16\,\sigma$, respectively. (d) Clear $100\,\%$ CH$_4$ (red), CO$_2$ (green), H$_2$O (orange), and Titan-like (purple) atmospheres cannot be rejected or confirmed.}
    \label{fig:tspec}
\end{figure*}

Given these short-timescale correlated noise structures, we performed the broadband and spectroscopic transit light curve analyses with five different systematics treatments, including Gaussian processes \citep[GPs,][]{rasmussen_gaussian_2006,ambikasaran_fast_2015} and detrending against the integrated flux in the H$\alpha$ stellar line (see Appendix~\ref{subapp:systematics}). All five systematics treatments resulted in statistically consistent transit spectra (Appendix~\ref{subapp:systematics}). The GP approach performed the best in terms of correlated noise removal, and therefore unless otherwise stated, quantitative results presented hereafter are derived from transit spectra resulting from this treatment (Figures~\ref{fig:tspec}(a)--\ref{fig:tspec}(b)). 

\section{Stellar Contamination and Planetary Atmosphere Fits} \label{sec:retrieval}

We investigated the contribution of stellar contamination from unocculted star spots and/or faculae \citep{sing_hubble_2011,mccullough_water_2014,rackham_access_2017,rackham_transit_2018,rackham_sag21_2023} to the transmission spectra through two different fitting approaches. The first approach fits a stellar contamination model to the transmission spectrum of each of the two visits. The best-fitting stellar contamination model for each visit is then divided out from each transmission spectrum, and the error-weighted average of the two stellar-contamination-corrected spectra is then subjected to a planetary atmosphere retrieval --- we term this approach a \textit{sequential} fit of the stellar contamination and planetary atmosphere (Section~\ref{subsec:sequential}). The second approach jointly fits a stellar contamination and planetary atmosphere model to each transit spectrum --- we term this a \textit{simultaneous} fit of the stellar contamination and planetary atmosphere (Section~\ref{subsec:simultaneous}). 

An alternative approach to treat stellar contamination is to fit the out-of-transit \textit{stellar} spectra to constrain the heterogeneity properties, and then use these constraints to correct the transmission spectrum \citep{wakeford_disentangling_2019,zhang_near-infrared_2018,garcia_hstwfc3_2022,rackham_towards_2023,berardo_empirically_2023}. This approach has the advantage of using information from the \textit{star} only to correct for stellar contamination, in contrast to the fit to the \textit{transmission spectrum}, which could be biased by the signal of a planetary atmosphere, thereby possibly erasing part of the planetary contribution during the stellar contamination correction. 

However, all of these approaches use theoretical stellar spectra such as the PHOENIX ACES models \citep{husser_new_2013}, which struggle to reproduce the observed, out-of-transit, spectrum of TRAPPIST-1, a late-type ultracool dwarf (see Appendix~\ref{app:stelspec}). These limitations in model fidelity impart an accuracy bottleneck in analyses of stellar impacts on transmission spectra and prevent us from determining the number of heterogeneities needed for contamination correction, resulting in error bars in the corrected transmission spectra that should actually be $\sim$1000--2000\,ppm per native-resolution pixel \citep[as seen in][and per the work in Appendix~\ref{app:stelspec}]{wakeford_disentangling_2019, garcia_hstwfc3_2022, rackham_towards_2023}, whether we fit the out-of-transit stellar spectra or the transmission spectra. These limitations also imply that the inferred spot and facula parameters likely do not reflect the true state of the heterogeneities at the time of the observations. This lack of model fidelity calls for more theoretical work and/or continuous observations of the star to characterize the heterogeneities empirically \citep[see, e.g.,][]{rackham_sag21_2023}. We did not apply the method based on out-of-transit spectra because for this particular approach the mismatch of the models to the data may lead to a stellar contamination correction that is largely inconsistent with the actual contribution of contamination to the transmission spectra. 

\subsection{Sequential Fit} \label{subsec:sequential}

The stellar contamination model assumes that the visible stellar disk has two unocculted regions, a spot and a facula, for a total of three components, including the quiescent photosphere. We opted for three components as it is the simplest model that allows one component hotter and one colder than the quiescent photosphere. The covering fractions of the spot and facula can go to zero, hence allowing a two- and a single-component scenario. We did not compare our results with models with more components, as the reduced $\chi^2$ values were already close to 1, i.e. the three-component model already provided a reasonable fit to the observations (Section~\ref{subsec:results_stelconam}). However, we do acknowledge that the transit depth precision may be underestimated because it does not fully account for the uncertainty on the number of components. More analysis, and likely more observations, will be needed to determine the number of components that best fit the data. 

We compute a weighted average of three PHOENIX ACES stellar spectra \citep{husser_new_2013} at the temperatures and gravities of each component to simulate the spectrum of a heterogeneous disk. The weight of each spectrum is determined by the size of the spot and of the facula. We then compute the expected transmission spectrum assuming an atmosphere-less planet transiting this heterogeneous disk \citep[Equation~3 from][]{rackham_transit_2018}. 

{
\begin{deluxetable*}{lcccccc}
    \tabletypesize{\scriptsize}
    \tablewidth{0pt} 
    \tablecaption{Priors and posteriors of the stellar contamination parameters from the sequential and simultaneous fits of the stellar contamination and planetary atmosphere.}
        \tablehead{
         \colhead{Stellar Contamination Parameter} & \multicolumn{2}{c}{Priors} & \multicolumn{4}{c}{Posteriors} \\
         & Sequential & Simultaneous & \multicolumn{2}{c}{Sequential} & \multicolumn{2}{c}{Simultaneous} \\
          & Visits 1 \& 2 & Visits 1 \& 2 & Visit 1 & Visit 2 & Visit 1 & Visit 2
         }
         \startdata
         Spot covering fraction, $f_{\rm spot}$ & $\mathcal{U}(0, 1)$ & $\mathcal{U}(0, 0.5)$ & $0.29_{-0.10}^{+0.11}$ & $0.16_{-0.08}^{+0.11}$ & $0.23_{-0.08}^{+0.10}$ & $0.24_{-0.14}^{+0.15}$ \\
         Facula covering fraction, $f_{\rm fac}$ & $\mathcal{U}(0, 1)$ & $\mathcal{U}(0, 0.5)$ & $0.05_{-0.04}^{+0.08}$ & $0.29_{-0.13}^{+0.12}$ & $0.07_{-0.04}^{+0.09}$ & $0.27_{-0.18}^{+0.15}$ \\
         Photospheric temperature, $T_{\rm phot}$ (K) & $\mathcal{N}(T_{\star,\,\rm eff}, \sigma_{T_{\star,\,\rm eff}})$ & $\mathcal{N}(T_{\star,\,\rm eff}, \sigma_{\rm T_{\star,\,\rm eff}})$ & $2567_{-46}^{+52}$ & $2557_{-45}^{+46}$ & $2571_{-42}^{+43}$ & $2576_{-37}^{+34}$ \\
         Stellar photosphere gravity, $\log{g_{\rm phot}}$ (dex) & 5.0 & $\mathcal{N}(\log{g_\star}, \sigma_{\log{g_\star}})$ & 5.0 & 5.0 & $5.24^{+0.01}_{-0.01}$ & $5.24^{+0.01}_{-0.01}$ \\
         Spot temperature, $T_{\rm spot}$ or $\Delta T_{\rm spot}$ (K) & $\mathcal{U}(-266, -100)$\tablenotemark{\rm\scriptsize a} & $\mathcal{U}(2300, T_{\star,\,\rm eff} + 3\,\sigma_{\rm T_{\star,\,\rm eff}})$\tablenotemark{\rm\scriptsize b} & $-197_{-62}^{+56}$\tablenotemark{\rm\scriptsize a} & $-178_{-60}^{+54}$\tablenotemark{\rm\scriptsize a} & $2371^{+70}_{-49}$\tablenotemark{\rm\scriptsize b} & $2431^{+83}_{-89}$\tablenotemark{\rm\scriptsize b} \\
         Facula temperature, $T_{\rm fac}$ or $\Delta T_{\rm fac}$ (K) & $\mathcal{U}(100, 1000)$\tablenotemark{\rm\scriptsize a} & $\mathcal{U}(T_{\star,\,\rm eff} - 3\,\sigma_{\rm T_{\star,\,\rm eff}}, 1.2\,T_{\star,\,\rm eff})$\tablenotemark{\rm\scriptsize b} & $+214_{-88}^{+257}$\tablenotemark{\rm\scriptsize a} & $+153_{-33}^{+61}$\tablenotemark{\rm\scriptsize a} & $2706^{+124}_{-86}$\tablenotemark{\rm\scriptsize b} & $2734^{+105}_{-53}$\tablenotemark{\rm\scriptsize b} \\
         Spot surface gravity, $\log{g}_{\rm spot}$ (dex) & \multirow{2}{*}{$\mathcal{U}(2.5, 5.5)$} & $\mathcal{U}(3.0, 5.4)$ & \multirow{2}{*}{$4.3_{-0.5}^{+0.2}$} & \multirow{2}{*}{$5.22_{-0.12}^{+0.11}$} & $4.55_{-0.42}^{+0.57}$ & $5.33_{-0.15}^{+0.05}$ \\
         Facula surface gravity, $\log{g}_{\rm fac}$ (dex) &  & $\mathcal{U}(3.0, 5.4)$ &  &  & $4.49_{-0.81}^{+0.57}$ & $5.28_{-0.25}^{+0.09}$ \\
         Planet radius $R_{\rm p}$ ($R_\oplus$) & ... & $\mathcal{U}(0.85, 1.15)\,R_{\rm p,\,obs}/R_\oplus$ & $1.09^{+0.01}_{-0.01}$ & $1.13^{+0.01}_{-0.01}$ & $1.10^{+0.01}_{-0.01}$ & $1.12^{+0.01}_{-0.01}$ \\
         Planet-to-star radius ratio, squared $(R_{\rm p}/R_\star)^2$ (ppm) & $\mathcal{U}(0.8, 1.2)\,{\rm med}(T_{\rm depth})$ &  ... & $6951_{-142}^{+144}$ & $7397_{-150}^{+148}$ & $7062_{-128}^{+128}$ & $7321_{-131}^{+131}$ \\
        \enddata
    \tablecomments{In the prior columns, $T_{\star,\,\rm eff}=2566\,$K, $\log{g_\star}=5.24$, and $R_{\rm p,\,obs}=1.116\,R_\oplus$ are adopted from the literature \citep{agol_refining_2021}, with slightly more conservative uncertainties $\sigma_{\rm T_{\star,\,\rm eff}}=50\,$K and $\sigma_{\log{g_\star}}=0.01$, and $\rm med(T_{depth})$ is the median of the measured transit depths of each visit. In the posterior columns, the leftmost value is the $50^{\rm th}$ percentile of the posterior distribution. Upper and lower uncertainties are computed from the $16^{\rm th}$, $50^{\rm th}$, and $84^{\rm th}$ percentiles.
    }
    \tablenotetext{\rm a}{Temperature difference with respect to the quiet photosphere.}
    \tablenotetext{\rm b}{Absolute temperature, \textit{not} relative to the quiet photosphere.}
    \label{tab:priors_posteriors_tls}
\end{deluxetable*}
}

Since the two visits were approximately $1.5\,$days apart \citep{gillon_seven_2017} and adopting a stellar rotation period of $3.3\,$days \citep{luger_seven-planet_2017}, the two visits were about half a stellar rotation apart, meaning that the spot and facula properties can be different between these two visits. We thus fitted the stellar contamination model to the transit spectrum of each visit separately, inferring the properties of the unocculted stellar heterogeneities, i.e. their temperature differences with respect to the quiescent photosphere, surface gravities, and covering fractions, with the priors listed in Table~\ref{tab:priors_posteriors_tls}. We forced the sum of both covering fractions to remain smaller than 1. Following Fournier-Tondreau et al. (2023, submitted), we set the surface gravity as a free parameter as a proxy of the pressure in the heterogeneity, which may be affected by the presence of magnetic fields \citep{solanki_sunspots_2003,bruno_hiding_2021}. For simplicity, we forced both heterogeneities to have the same surface gravity in this particular fit. We also set the planet-to-star radius ratio and the photospheric temperature as free parameters to account for their uncertainties. 

We then divided out the median stellar contamination model from each of the two transmission spectra and multiplied back the median squared planet-to-star radius ratio. We took an error-weighted average of the two stellar-contamination-corrected transmission spectra, yielding a stellar-contamination-corrected and combined transmission spectrum.

Before performing the planetary atmosphere retrieval, we first compared the stellar-contamination-corrected and combined transmission spectrum to a selection of SCARLET planetary atmosphere forward models \citep{benneke_atmospheric_2012,benneke_how_2013,benneke_strict_2015,benneke_water_2019,benneke_sub-neptune_2019,pelletier_where_2021} of TRAPPIST-1~b. We computed the $\chi^2$ statistic between each atmospheric model and the corrected and combined spectrum, and measured how far out in the distribution tail the statistic fell, in units of the standard deviation of the $\chi^2$ distribution \citep{gregory_bayesian_2005}. This frequentist calculation provides us with rejection (or detection) confidence levels for specific planetary atmosphere scenarios, which can guide the atmospheric retrieval.

\begin{figure*}
    \centering
    \includegraphics[width=.7\textwidth]{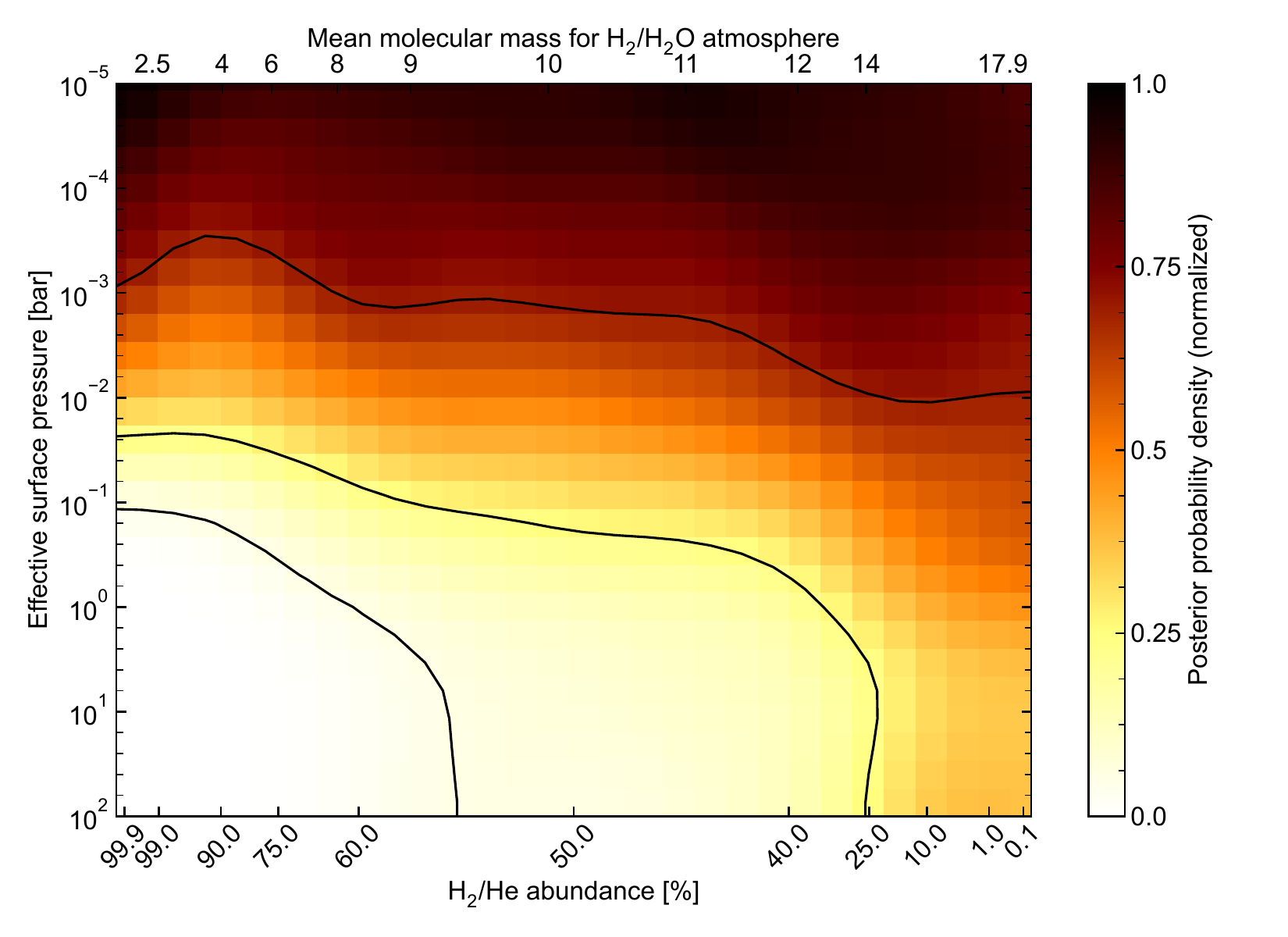}
    \caption{Joint constraints on the H$_2$/He abundance and the atmospheric pressure on TRAPPIST-1~b resulting from the sequential fit of the stellar contamination and planetary atmosphere. The color shading illustrates the posterior probability density, darker colors corresponding to higher probabilities. Contours indicate the 1-$\sigma$, 2-$\sigma$, and 3-$\sigma$ Bayesian credible regions. The displayed posterior probability is marginalized over the H$_2$O, CH$_4$, CO, CO$_2$, NH$_3$, and N$_2$ abundances. Hydrogen-dominated scenarios with high-altitude clouds are at the top left corner of the plot and cloud-free, volatile-rich, high-mean-molecular-mass atmospheres are at the bottom right. Any hydrogen-rich atmospheres without high-altitude clouds, at the bottom left, are robustly ruled out. The second horizontal axis at the top shows the mean molecular mass representative for a pure H$_2$/H$_2$O atmosphere.}
    \label{fig:AtmosphericRetrievalMainFig}
\end{figure*}

We used SCARLET to perform a planetary atmosphere retrieval on the stellar-contamination-corrected and combined transmission spectrum. Since the frequentist comparison between atmospheric forward models and the corrected and combined spectrum did not detect an atmosphere, this retrieval analysis explored the range of mean molecular masses and surface/cloud pressures that remain consistent with the data, similar to an analysis done with GJ~1214~b and GJ~436~b \citep{benneke_distinguish_2013,kreidberg_clouds_2014,knutson_featureless_2014}. We used a parameter space that efficiently samples the full range of models between hydrogen-dominated and volatile-dominated atmospheres. We adopted a mix of H$_2$/He at solar ratio, along with molecules of higher mean molecular mass, namely H$_2$O, CH$_4$, CO, CO$_2$, NH$_3$, and N$_2$. This is because TRAPPIST-1 planets could host a wide variety of secondary atmospheres with outgassing/volcanism and cometary delivery as their main sources \citep[][and references therein]{turbet_review_2020}, along with photochemistry shaping the composition of these atmospheres given the strong X and UV radiation and flares emitted by the host star \citep{wheatley_strong_2017,bourrier_temporal_2017,vida_frequent_2017}. Optically thick and grey clouds were also included in the model, such that the pressure parameter corresponds to the pressure either at the ground or at the top of the cloud. We parameterized the forward model to sample the pressure and H$_2$/He abundance relative to all gases, using the two-component centered-log-ratio transform \citep{benneke_how_2013}, also illustrated by the horizontal axis stretch of Figure~\ref{fig:AtmosphericRetrievalMainFig}. 

\subsection{Simultaneous Fit} \label{subsec:simultaneous}

For the simultaneous fit of the stellar contamination and planetary atmosphere, we used \texttt{POSEIDON} \citep{macdonald_hd_2017,macdonald_poseidon_2023} to analyze the transmission spectrum of each visit individually. We parameterized stellar contamination via a two-heterogeneity model defined by 8 free parameters (Fournier-Tondreau et al. 2023, submitted; priors listed in Table~\ref{tab:priors_posteriors_tls}): the spot and facula coverage fractions, temperatures, and local surface gravities, as well as the stellar photosphere temperature and surface gravity. We calculated the impact of stellar contamination by interpolating PHOENIX models \citep{husser_new_2013} using the \texttt{MSG} package \citep{townsend_msg_2023}. We fitted for the planetary atmosphere via a 6-parameter model: the atmosphere radius at the 1\,bar reference pressure, $R_{\rm p,\,ref}$ \citep[with prior $\mathcal{U}(0.85, 1.15)\,R_{\rm p,\,obs}$, where $R_{\rm p,\,obs} = 1.116\,R_{\oplus}$;][]{agol_refining_2021}, the atmospheric temperature, $T$ ($\mathcal{U}(200, 900)$\,K), and the H$_2$, H$_2$O, and CO$_2$ mixing ratios, $X_{i}$ \citep[centered log-ratio priors from $10^{-12}$ to 1 with the remainder of the atmosphere filled with N$_2$;][]{Lustig-YaegerFu2023}. We calculated each model transmission spectrum on a wavelength grid at a resolving power of $R =$ 20,000 from 0.58 to 3.0\,$\mu$m, including both the planetary atmosphere and stellar contamination contribution.

\section{Results and Discussion} \label{sec:results}

\subsection{Evidence for Stellar Contamination} \label{subsec:results_stelconam}

The transmission spectra of TRAPPIST-1~b notably differ between the two visits (Figures~\ref{fig:tspec}(a)--\ref{fig:tspec}(b)), with different wavelength-dependent slopes, spanning a few hundreds of ppm, and different absorption band-like features. These differences can be explained by stellar contamination and its variation between the two epochs. 

The stellar contamination fit of the sequential fit shows that the transmission spectrum of the first visit is explained by an unocculted spot (with no evidence for faculae), with a covering fraction of $29^{+11}_{-10}\,\%$ and a temperature difference with respect to the photosphere of $-197^{+56}_{-62}$\,K (Figure~\ref{fig:tspec}(a) and Table~\ref{tab:priors_posteriors_tls}). The second visit shows a preference for an unocculted facula (with no evidence for spots), causing the strong slope in the blue part of the spectrum (Figure~\ref{fig:tspec}(b)). The covering fraction of this facula is $29^{+12}_{-13}\,\%$ and its temperature difference is $+153^{+61}_{-33}\,$K. The stellar contamination model cannot \textit{perfectly} fit the observed spectra, which is explained by the mismatch between the stellar models and the observations (see Appendix~\ref{app:stelspec}), but it can, to some extent, explain the slope bluewards of $1.3\,\mu$m as well as some features near $1.4\,\mu$m, especially for the second visit. 

\begin{figure*}
    \centering
    \includegraphics[width=\textwidth]{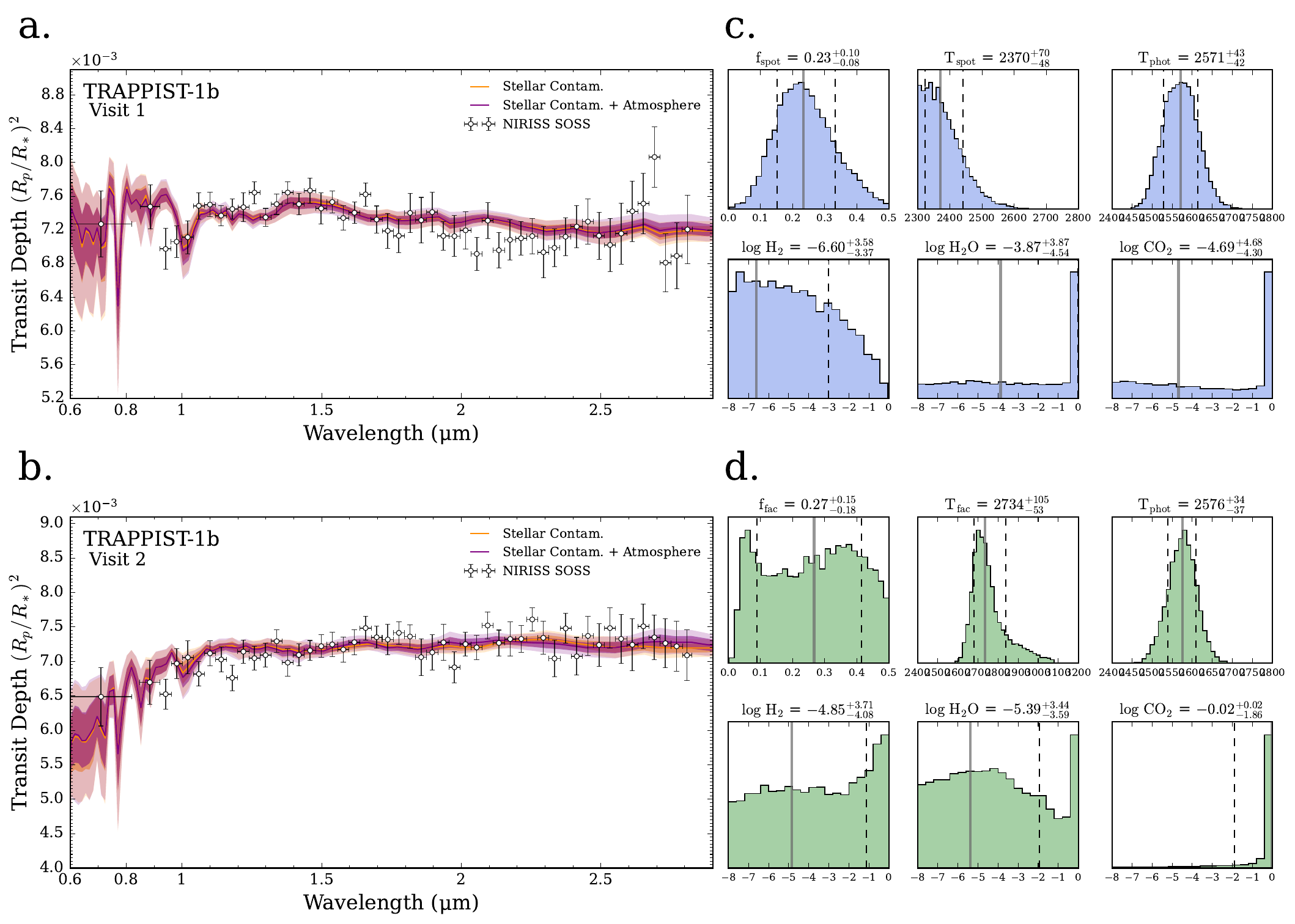}
    \caption{Atmospheric and stellar properties from the transmission spectrum of TRAPPIST-1~b from the simultaneous stellar contamination and planetary atmosphere fit. (a)--(b) Retrieved model spectrum of TRAPPIST-1~b from two retrievals: a stellar contamination model with unocculted spots and faculae (orange) and a joint stellar contamination and planetary atmosphere model (purple), for visits 1 and 2. The median retrieved model (solid lines) and the corresponding 1-$\sigma$ and 2-$\sigma$ confidence intervals (dark and light contours) are compared to the observed TRAPPIST-1~b spectrum (black points). Error bars are the 1-$\sigma$ uncertainties. (c)--(d) Corresponding posterior probability distributions from the joint stellar contamination and planetary atmosphere retrieval for visits 1 and 2. The TRAPPIST-1~b spectra are well fitted by the stellar contamination model ($\chi^2_{\nu}$ = 1.07 and 0.75 for the best-fitting model spectra to visit 1 and 2, respectively), with no fit improvement from adding a thick planetary atmosphere. However, the CO$_2$ and H$_2$O posteriors indicate that a high-mean-molecular-mass atmosphere remains consistent with the TRAPPIST-1~b observations.
    }
    \label{fig:retrieval_onestep}
\end{figure*}

The results from the simultaneous stellar contamination and planetary atmosphere fit (Figure~\ref{fig:retrieval_onestep} and Appendix~\ref{app:corner}) are consistent with the unocculted spot and faculae properties inferred from the stellar contamination fit of the sequential fit (Table~\ref{tab:priors_posteriors_tls}). The first visit is explained by unocculted spots about 200\,K cooler than the stellar photosphere, whilst the second visit requires unocculted faculae about 160\,K hotter than the photosphere (Figures~\ref{fig:retrieval_onestep}(c)--\ref{fig:retrieval_onestep}(d)). 
The stellar contamination model is moderately to strongly preferred over a flat spectrum model, with log Bayes factors of 5.3 and 2.9 for the first and second visits, respectively, following Table~1 from \citet{trotta_bayes_2008}. 
Comparing the transmission spectra from the two visits to a stellar-contamination-only model and a joint stellar contamination and planet atmosphere model (Figures~\ref{fig:retrieval_onestep}(a)--\ref{fig:retrieval_onestep}(b)) shows that the observations can be described by stellar contamination alone; there is no significant improvement from adding an atmosphere. 
The $\chi^2$ from the fits are as follows: for visit 1, $\chi^2$ = 44.0 (with 41 degrees of freedom) for a retrieval with only stellar contamination and 43.5 (with 37 degrees of freedom) for stellar contamination and a planetary atmosphere. For visit 2, $\chi^2$ = 30.6 and 29.6 for the same models. 

The inferred stellar heterogeneity parameters and atmospheric constraints from the simultaneous fit are generally similar to those inferred from the sequential fit (Section~\ref{subsec:results_atm}), likely due to the lack of features from the planetary atmosphere. These two analyses of the NIRISS/SOSS observations strongly support previous evidence that stellar contamination is a critical consideration when interpreting any transmission spectrum of planets transiting active stars \citep[e.g.,][]{moran_high_2023}, in particular TRAPPIST-1 planets \citep{zhang_near-infrared_2018,wakeford_disentangling_2019,ducrot_trappist-1_2020,garcia_hstwfc3_2022}. It is intriguing though that the amplitude of the TLS effect is over an order of magnitude larger in the two TRAPPIST-1~b spectra ($\sim$750--1000\,ppm peak-to-peak) than in the TRAPPIST-1~g spectra obtained with NIRSpec/Prism over two visits (Benneke et al., in preparation), also part of the JWST TRAPPIST-1 atmospheric reconnaissance program. This may indicate that the amplitude of the TLS effect varies strongly with epoch and/or transit impact parameter \citep[$b\sim0.1$ and $b\sim0.4$ for TRAPPIST-1~b and g, respectively;][]{agol_refining_2021}, but this remains speculative given the small number of observations. All this underscores the need for more observations to better quantify the impact of stellar contamination at any given epoch and geometrical configuration. 

\subsection{Constraints on the Planetary Atmosphere} \label{subsec:results_atm}

The median error bar of the stellar-contamination-corrected and combined transmission spectrum at $R \sim 15$ is $89\,$ppm, not accounting for the ${\sim}1800$-ppm per-pixel model-driven uncertainty associated with the lack of stellar model fidelity (see Appendix~\ref{app:stelspec}). Signatures of a cloud-free atmosphere on TRAPPIST-1~b could have amplitudes of $100$--$200\,$ppm, depending on the exact atmospheric conditions \citep{lincowski_evolved_2018}. The precision achieved with two NIRISS/SOSS transits is thus insufficient to confidently detect or reject such atmospheres yet, but it shows that this objective could be within reach with a reasonable number of visits considering the expected lifetime of JWST and its instrument.

The frequentist comparison between atmospheric forward models and the stellar-contamination-corrected and combined transmission spectrum rejects cloud-free, hydrogen-rich atmospheres at 1$\times$ and 100$\times$\,solar metallicity at 29 and 16\,$\sigma$, respectively, further confirming previous analyses of \textit{HST} observations \citep{de_wit_combined_2016}. We also compared the transmission spectrum to a limited set of cloud-free secondary atmospheres, namely pure CH$_4$, H$_2$O, and CO$_2$ models, as well as a clear Titan-like scenario (Figure~\ref{fig:tspec}(d)), but none of these scenarios can be confidently confirmed or rejected at present (0.82, 0.48, 0.84, and 0.44\,$\sigma$, respectively). We conclude that the current NIRISS transmission spectrum (2 visits), after stellar contamination correction, does not show any signatures of an atmosphere.

The sequential fit of stellar contamination and planetary atmosphere finds that the TLS-corrected and combined transmission spectrum is only consistent at $3\,\sigma$ with a particular range of atmospheres (Figure~\ref{fig:AtmosphericRetrievalMainFig}), from clear atmospheres with a mean molecular mass greater than approximately 9 a.m.u. to low-mean-molecular-mass atmospheres with cloud surface pressures less than $0.1\,$bar. However, if TRAPPIST-1~b has retained a primordial atmosphere, cloud formation could be unlikely given the amount of radiation the planet receives \citep{morley_thermal_2015,de_wit_combined_2016}. Clouds are also unlikely, at least on the dayside of TRAPPIST-1~b, given the near-zero Bond albedo inferred from the JWST/MIRI secondary eclipse observations \citep{greene_thermal_2023}. These Bayesian results are consistent with the frequentist rejection of the clear hydrogen-rich forward models. 

The simultaneous fit of stellar contamination and planetary atmosphere finds a 3-$\sigma$ upper limit on the H$_2$ abundance of $52\,\%$ from the first visit (Figure~\ref{fig:retrieval_onestep}(c) and Appendix~\ref{app:corner}), consistent with the results from the sequential fit. The second visit alone is inconclusive on the H$_2$ abundance, as the faculae-dominated spectrum yields a bimodal posterior distribution, allowing an unphysical atmosphere with $100\,\%$ H$_2$ and no trace volatiles (Figure~\ref{fig:retrieval_onestep}(d) and Appendix~\ref{app:corner}). The peaks in the posteriors for H$_2$O and CO$_2$ are manifestations of the uninformative centered log-ratio prior that treats all the gases equally (i.e. if there is an atmosphere with a high mean molecular mass, either or both H$_2$O and CO$_2$ could be the background gas). The results from this simultaneous retrieval of the planetary atmosphere and stellar contamination remain consistent with a clear, high-mean-molecular-mass atmosphere dominated by N$_2$, H$_2$O, and/or CO$_2$ (see the H$_2$O and CO$_2$ histogram peaks in Figures~\ref{fig:retrieval_onestep}(c)--\ref{fig:retrieval_onestep}(d)).

In addition to being consistent with pre-JWST observations, the conclusions from both the sequential and simultaneous fits of stellar contamination and planetary atmosphere are consistent with the conclusions from the JWST/MIRI secondary eclipse observations \citep{greene_thermal_2023,ih_constraining_2023} in that the transits reveal no evidence for an atmosphere. Both transit and secondary eclipse observations agree that TRAPPIST-1~b does not have a thick atmosphere, at least not one with a low mean molecular mass based on the transit observations. The MIRI observations provide additional constraints by rejecting certain atmospheres containing CO$_2$, as well as atmospheres containing no CO$_2$ for certain pressures and compositions \citep[right panel of Figure~2 from][]{ih_constraining_2023}, further ruling out regions of the parameter space allowed by the transmission data. In this sense, the transmission spectra are less constraining in terms of atmospheric characterization for TRAPPIST-1~b, but they partially confirm the emission data and, more importantly, they bring insight to the stellar contamination aspect of transmission spectroscopy, which is currently the preferred technique to probe the potential atmospheres of the outer, cooler TRAPPIST-1 planets.

Being closest to the host star, TRAPPIST-1~b is the planet in this system most likely to have lost its atmosphere through atmospheric escape processes. Because of the highly irradiated environment that TRAPPIST-1~b is subjected to, the potential absence of an atmosphere on TRAPPIST-1~b would not necessarily forbid atmospheres on the other planets in the system, especially the outer planets \citep{krissansen-totton_implications_2023}. Moreover, the non-detection of an atmosphere on TRAPPIST-1~b can provide insight into the evolutionary history of the system. Results from this work and from the eclipse observations \citep{greene_thermal_2023,ih_constraining_2023} should thus not discourage from monitoring transits of the TRAPPIST-1 planets to search for atmospheric signatures.

\section{Conclusion} \label{sec:conclusion}

We presented the first transmission spectra of the TRAPPIST-1~b planet with NIRISS/SOSS. Stellar contamination, whose manifestation differs from one visit to the other, dominates the transmission spectra at the level of several hundreds of ppm and must be properly disentangled from planetary signals. Accounting for stellar contamination, our analyses confidently reject clear H$_2$-rich atmospheres for TRAPPIST-1~b. However, the current SOSS observations cannot confirm whether TRAPPIST-1~b is a bare rock or if it has a thin and/or high-mean-molecular-mass atmosphere, although the latter scenario has been rejected for certain compositions by secondary eclipse observations with JWST/MIRI \citep{greene_thermal_2023,ih_constraining_2023}. As TRAPPIST-1~b is the planet in the system most likely to have lost its atmosphere \citep{krissansen-totton_implications_2023}, a lack of atmosphere would not suggest that the outer planets are bare rocks and thus should not discourage future observations of the TRAPPIST-1 system. Assessing the presence of an atmosphere on the habitable-zone and outer TRAPPIST-1 planets is currently only possible with transmission spectroscopy. Given the lack of stellar model fidelity, additional theoretical work \citep[e.g.,][]{witzke_faculae_2022, rackham_sag21_2023} and/or observations of the host star \citep[e.g.,][]{berardo_empirically_2023} are necessary to provide better constraints on the contribution of stellar contamination to future transmission spectra.\\


This work is based on observations made with the NASA/ESA/CSA James Webb Space Telescope. The data were obtained from the Mikulski Archive for Space Telescopes at the Space Telescope Science Institute, which is operated by the Association of Universities for Research in Astronomy, Inc., under NASA contract NAS 5-03127 for JWST. These observations are associated with program \#2589.
This research has made use of the VizieR catalogue access tool, CDS, Strasbourg, France (DOI : 10.26093/cds/vizier). Specifically, VizieR catalogue J/A+A/640/A112 (Ducrot E.) \citep{vizier_ducrot_trappist-1_2020} was used to produce Figure~\ref{fig:tspec}(c). The original description of the VizieR service was published in 2000, A\&AS 143, 23.
O.L. acknowledges financial support from the Fonds de recherche du Qu\'{e}bec --- Nature et technologies (FRQNT), and funding from the Trottier Family Foundation in their support of iREx. 
B.B. acknowledges financial support from the Natural Sciences and Engineering Research Council (NSERC) of Canada and the FRQNT. 
R.D. acknowledges financial support from the NSERC and the Canadian Space Agency through grants number 22JWG01-3 and 22EXPJWST. 
R.J.M. acknowledges support for this work provided by NASA through the NASA Hubble Fellowship grant HST-HF2-51513.001 awarded by the Space Telescope Science Institute, which is operated by the Association of Universities for Research in Astronomy, Inc., for NASA, under contract NAS5-26555.
C.P. acknowledges financial support from the FRQNT, the Technologies for Exo-Planetary Science (TEPS) Trainee Program and the NSERC Vanier Scholarship. 
A.L'H. and M.F.-T. acknowledge financial support from the FRQNT. 
B.V.R thanks the Heising-Simons Foundation for support. This material is based upon work supported by the National Aeronautics and Space Administration under Agreement No. 80NSSC21K0593 for the program ``Alien Earths''. The results reported herein benefited from collaborations and/or information exchange within NASA's Nexus for Exoplanet System Science (NExSS) research coordination network sponsored by NASA's Science Mission Directorate.
L.D. acknowledges support from the Banting Postdoctoral Fellowship program, administered by the Government of Canada and the L'Oréal-UNESCO For Women in Science program.
The authors wish to thank Kevin Volk for sharing the reference file required to flux-calibrate the SOSS spectra of TRAPPIST-1.

%

\vspace{5mm}
\facilities{JWST(NIRISS)}


\software{\texttt{supreme-SPOON}\footnote{\url{https://github.com/radicamc/supreme-spoon}} \citep{feinstein_early_2022,radica_applesoss_2022,radica_awesome_2023},
          \texttt{ATOCA}\footnote{\url{https://github.com/AntoineDarveau/atoca_demo}} \citep{darveau-bernier_atoca_2022},
          \texttt{NAMELESS} \citep{feinstein_early_2022,coulombe_broadband_2023},
          \texttt{jwst}\footnote{\url{https://github.com/spacetelescope/jwst}},
          \texttt{emcee}\footnote{\url{https://github.com/dfm/emcee})} \citep{foreman-mackey_emcee_2013},
          \texttt{corner}\footnote{\url{https://github.com/dfm/corner.py}} \citep{foreman-mackey_cornerpy_2016},
          \texttt{batman}\footnote{\url{https://github.com/lkreidberg/batman}} \citep{kreidberg_batman_2015},
          \texttt{POSEIDON}\footnote{\url{https://github.com/MartianColonist/POSEIDON}} \citep{macdonald_hd_2017,macdonald_poseidon_2023},
          PHOENIX \citep{husser_new_2013},
          SPHINX \citep{iyer_sphinx_2023,iyer_sphinx_2023_7416042},
          \texttt{MSG}\footnote{\url{https://github.com/rhdtownsend/msg}} \citep{townsend_msg_2023},
          \texttt{PyMultiNest}\footnote{\url{https://github.com/JohannesBuchner/PyMultiNest}} \citep{Feroz2009,Buchner2014},
          \texttt{astropy}\footnote{\url{https://www.astropy.org/}} \citep{the_astropy_collaboration_astropy_2013,the_astropy_collaboration_astropy_2018},
          \texttt{numpy}\footnote{\url{https://github.com/numpy/numpy}} \citep{harris_array_2020}, 
          \texttt{matplotlib}\footnote{\url{https://github.com/matplotlib/matplotlib}} \citep{hunter_matplotlib_2007}, \texttt{scipy}\footnote{\url{https://github.com/scipy/scipy}} \citep{virtanen_scipy_2020}
          }



\appendix

\section{Data reduction} \label{app:reduction}

\subsection{\texttt{supreme-SPOON}} \label{subapp:supreme-spoon}

The \texttt{supreme-SPOON} pipeline has already been presented in detail in \citet{radica_awesome_2023} \citep[see also][]{feinstein_early_2022, coulombe_broadband_2023}, and the steps we follow here closely mirror those other analyses. We process the data of both visits through the standard \texttt{supreme-SPOON} stage 1, which includes a group-level correction of 1/$f$ noise, and stage 2. During background correction, we found that the SOSS SUBSTRIP256 background model provided by STScI could not completely remove the pick-off mirror jump in the zodiacal background near spectral pixel 700 in Figure~\ref{fig:reduction}(b). We therefore constructed a new background model from a high-signal-to-noise stacked image of each TRAPPIST-1 time series. We first masked all traces (target and field contaminants) and bad pixels. Then, assuming each column sees a constant background level, we adopted the 10$^{\rm th}$ percentile value of each column and low-pass-filtered these values to produce a smoothed background. We performed these steps to build the background model with the deep stack rotated such that the background discontinuity at $x \approx 700$ was perfectly vertical. The model was then reprojected in 2D, de-rotated and subtracted from the time series. The 1D spectral extraction was performed with a 25-pixel-wide box aperture. Dilution introduced due to the overlap of the first and second spectral orders on the detector is predicted to be negligible for relative flux measurements \citep[e.g., transmission spectra,][]{darveau-bernier_atoca_2022,radica_applesoss_2022}: we estimate that the amount of dilution introduced by the order overlap would be less than $5\,$ppm throughout the $0.6$--$2.8\,\mu$m wavelength range \citep[following Equation~5 from][]{darveau-bernier_atoca_2022}, significantly smaller than the transit depth precision.

\subsection{\texttt{NAMELESS}} \label{subapp:nameless}

\begin{figure*}
    \centering
    \includegraphics[width=.8\textwidth]{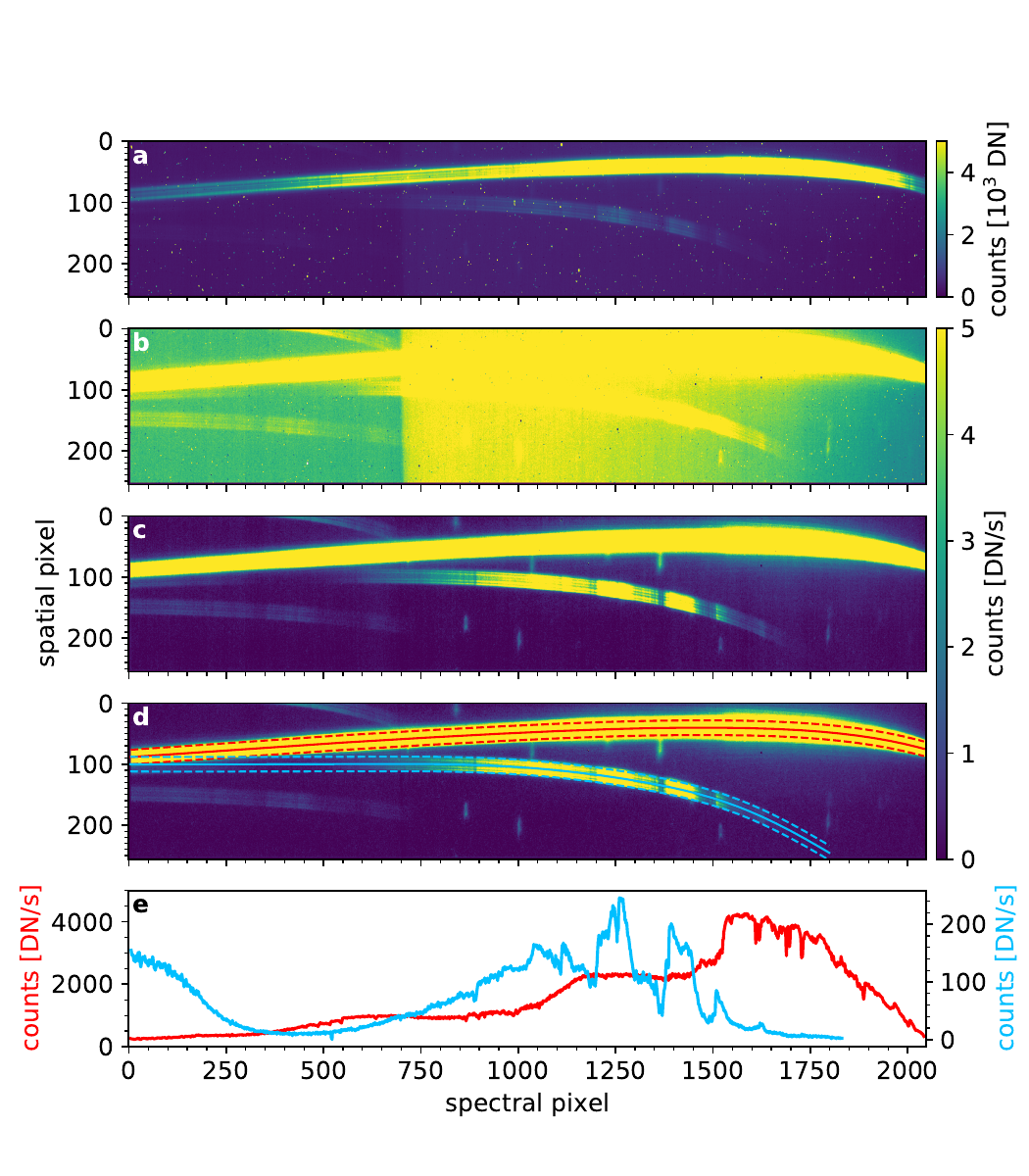}
    \caption{Reduction and spectral extraction of two SOSS spectral orders with the \texttt{NAMELESS} pipeline. (a) Last group of the 10$^{\rm th}$ integration of visit 1, after super-bias subtraction and non-linearity correction. (b) Detector image of the 10$^{\rm th}$ integration after ramp-fitting of the 18 groups comprising each integration. (c) Frame after bad pixel and outlier correction, as well as background subtraction. (d) Frame after correction of the 1/$f$ noise. The center (full line) and edges (dashed lines) of the fixed aperture boxes used to extract the first (red) and second (blue) spectral orders are shown. (e) Extracted spectra of the first (red) and second (blue) spectral orders. The sharp rise in the spectrum of the second order when going to lower spectral pixels is caused by the overlap with the first order.}
    \label{fig:reduction}
\end{figure*}

We also reduced the data with the \texttt{NAMELESS} pipeline \citep{feinstein_early_2022,coulombe_broadband_2023,radica_awesome_2023} (Figure~\ref{fig:reduction}), which first runs all the steps of the \texttt{jwst} pipeline stage 1 except for the dark current subtraction, as it left residual 1/$f$ noise. A parallel reduction was also performed \textit{with} the dark current subtraction, but it had virtually no effect on the transmission spectrum. We then went through the \texttt{assign\_wcs}, \texttt{srctype}, and \texttt{flat\_field} steps of the \texttt{jwst} pipeline stage 2 before performing custom routines for the correction of bad pixels, background, as well as 1/$f$ noise \citep{coulombe_broadband_2023}.

As with the \texttt{supreme-SPOON} pipeline, we found that the 2D background model from STScI could not reproduce the observed background jump near spectral pixel 700. To address this issue and to avoid introducing dilution through background correction, we split the background model into two separate sections along the line tracing the jump in background flux and fitted their amplitude separately. The background model made from the two distinct sections was then subtracted from all integrations. Spectral extraction of the first and second spectral orders was performed with a 24-pixel-wide box aperture. 

\subsection{\texttt{SOSSISSE}} \label{subapp:sossisse}

As a third reduction pipeline, we used the SOSS-Inspired SpectroScopic Extraction (\texttt{SOSSISSE}) tool. The \texttt{SOSSISSE} tool uses rateints files, i.e., the post-ramp-fitting time series of 2D images of the trace, to express any given trace image in the time series as the linear combination of a reference trace and its spatial derivatives. This approach assumes that the trace will only suffer achromatic transforms and provides a residual map from which spectroscopic information is then retrieved. Small signals from planetary transits are thus encoded into residual maps that have a much smaller dynamic range than the input dataset, such that low-amplitude effects (e.g. 1/$f$ noise) are more readily handled.

The \texttt{SOSSISSE} tool first constructs a model that consists of 1) a high-signal-to-noise (high-SNR), normalized, out-of-transit, median trace, $M_{i, j}$, where $i$ and $j$ are the indices of the pixels in the $x$ (dispersion) and $y$ (cross-dispersion) directions, respectively, and 2) time-dependent morphological changes in the trace. These morphological changes are represented by spatial derivatives of $M_{i, j}$: shifts in the $x$ and $y$ directions ($\partial M / \partial x$ and $\partial M / \partial y$), a trace rotation ($\partial M / \partial \theta$), and changes in the contrast of the point spread function (PSF) structures expressed as the second derivative along the $y$ direction ($\partial^2 M / \partial y^2$). The model can be written as
\begin{equation}\label{eq:sossisse}
\begin{aligned}
    F_{i, j}(t) =& \ a(t) \, M_{i, j} + b(t) \, \frac{\partial M_{i, j}}{\partial x} + c(t) \, \frac{\partial M_{i, j}}{\partial y} \\ 
    & + d(t) \, \frac{\partial M_{i, j}}{\partial \theta} + e(t) \, \frac{\partial^2 M_{i, j}}{\partial y^2} \,,
\end{aligned}
\end{equation}
where $F_{i, j}(t)$ is the model trace flux at time $t$ on pixel $(i, j)$, and $a$, $b$, $c$, $d$, and $e$ are time-dependent scalars (Figures~\ref{fig:wlc_fit}(e)--\ref{fig:wlc_fit}(h) and \ref{fig:wlc_fit}(m)--\ref{fig:wlc_fit}(p)), fitted in a least-squares fashion to each 2D trace image. Scalars $b$, $c$, $d$, and $e$ can be interpreted as the amplitudes of their respective derivatives (e.g., $b(t)$ can be seen as ``$\mathrm{d}x$'', the amplitude of the trace displacement in the $x$ direction). 

These morphological changes in the trace should be accounted for to accurately measure the photometric signal, and the \texttt{SOSSISSE} tool includes them as part of the photometric measurement instead of attempting to remove them through detrending in a subsequent step. Indeed, the fitted amplitude $a(t)$ of the high-SNR trace consistently accounts for these morphological changes in the trace, ensuring that the photometric measurements are consistent across the entire time series. We also found that $e(t)$ is particularly effective at detecting telescope tilt events \citep{rigby_science_2023,doyon_near_2023}. These events can affect the shape of the trace, but may only slightly shift its overall position. Including this term in the model accurately captures the impact of any potential tilt events on the data. 

Once an optimal trace model has been found for each 2D image of the time series, \texttt{SOSSISSE} subtracts this best-fit trace model from each image, leaving only chromatic temporal variations, or ``residuals''. \texttt{SOSSISSE} then collapses each 2D residual image into a 1D spectrum by performing a least-square adjustment of the model trace $F$ at spectral pixel $(i, j)$ onto the residual. The underlying assumption is that residuals resulting from spectroscopic features will have the same morphology as the underlying trace. This least-square fitting is performed with the propagation of the uncertainty from the rateints files. We also account for the probability that per-pixel residuals are either consistent with their uncertainty or with a cosmic ray event by giving to each pixel a weight corresponding to 1 minus its likelihood of being affected by a cosmic ray. The time series $a(t)$, which can be interpreted as the broadband light curve, is finally added back to each spectral channel, yielding a spectral time series that can be used to perform the transit light curve fit (Section~\ref{sec:lcfit}).

\subsection{Consistency Across Reduction Pipelines} \label{subapp:reduction_consistency}

Similar structures are present across the spectral time series of the three pipelines. For example, the temporal slope in visit 1 and the same correlated noise pattern associated with stellar activity near the transit egress in visit 2 are present in all three reductions. There are some differences between reductions, typically localized in time and wavelength, but these are expected given the differences between the pipelines. 

\begin{figure*}
    \centering
    \includegraphics[width=1.0\textwidth]{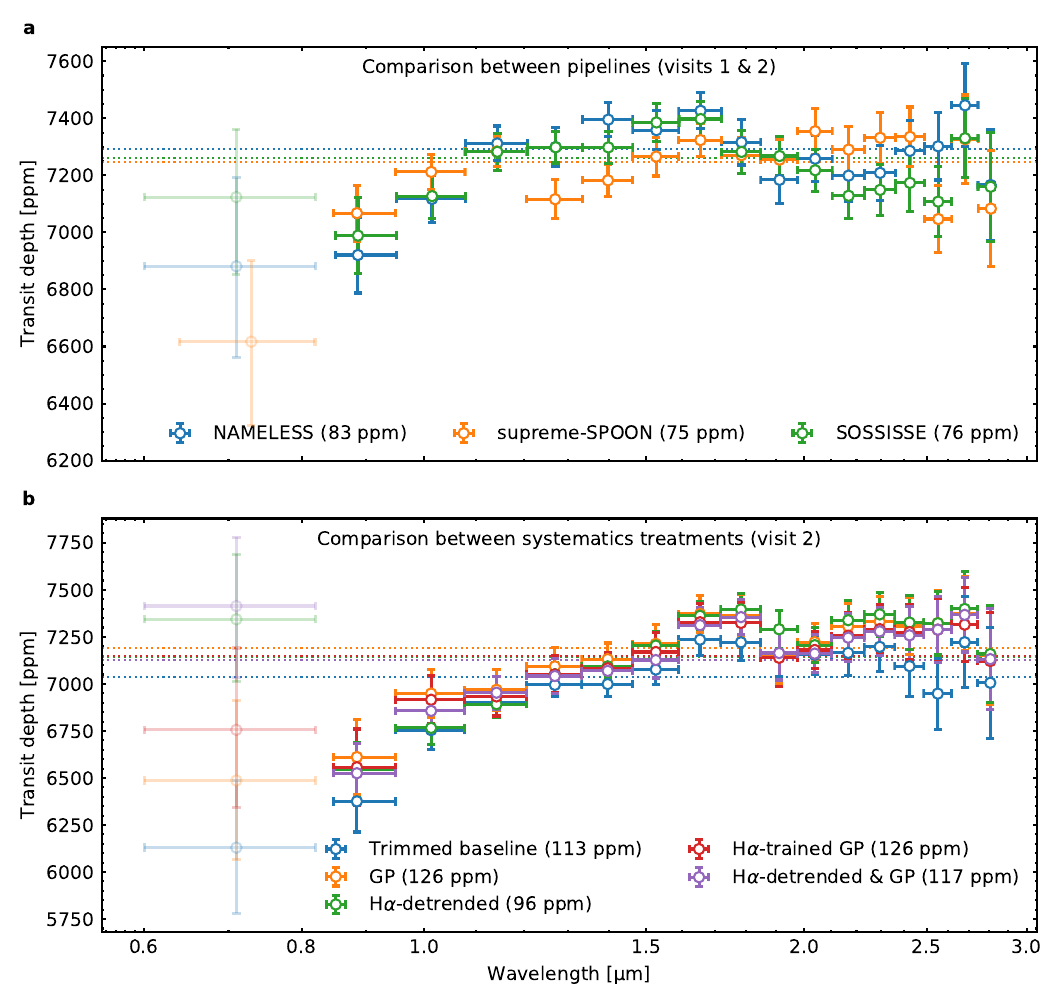}
    \caption{Comparison between reductions and systematics treatments. (a) Comparison between the (combined) transmission spectra resulting from the three reduction pipelines. Solid and semi-transparent points are from spectral orders 1 and 2, respectively. Horizontal dashed lines show the error-weighted average transit depths. (b) Comparison between the transit spectra of visit 2 resulting from the five different systematics treatments. In both panels, the numbers in parentheses in the legends are the median error bar across each spectrum. All vertical error bars are the 1-$\sigma$ uncertainties. All horizontal error bars represent the extent of each spectral bin.}
    \label{fig:tspec_pipelines_systematics}
\end{figure*}

We fitted the broadband and spectroscopic light curves of all three reductions to compare the resulting transmission spectra (Figure~\ref{fig:tspec_pipelines_systematics}(a)). Although the transit depths are generally consistent within their uncertainties, there are differences at the 100--200-ppm level in some wavelength bins. Differences of this order of magnitude between reductions have already been reported in JWST data \citep{feinstein_early_2022,coulombe_broadband_2023,rustamkulov_early_2022,radica_awesome_2023}, but these become more problematic for Earth-sized planets like TRAPPIST-1~b, whose expected atmospheric signatures have amplitudes of 100--200\,ppm \citep{lincowski_evolved_2018} at most. 

To assess the impact of the differences between reductions, we ran the simultaneous fit of the stellar contamination and planetary atmosphere (Section~\ref{subsec:simultaneous}) on all three sets of TRAPPIST-1~b transmission spectra resulting from the three pipelines, generally finding consistent conclusions. Regardless of the reduction, we see evidence of stellar contamination and cloud-free, hydrogen- and helium-rich models can be ruled out at high confidence. Secondary atmospheres, especially those with a low surface pressure, and/or clouds cannot be confidently rejected from these SOSS observations. Since the conclusions remain the same across all three reductions, we consider that our reductions are reliable for the purpose of this work. 

\section{Transit light curve fitting} \label{app:lcfit}

\subsection{Light Curve Preparation} \label{subapp:lcprep}

We started the transit light curve fit with a set of approximately 2048 light curves, each light curve corresponding to a spectral channel. For spectral order 1, we produced the broadband light curve by binning all light curves into a single bin, whereas the spectroscopic light curves were generated by binning the light curves down to 50 bins, equally spaced in wavelength. For spectral order 2, we first removed all light curves at wavelengths above $0.82\,\mu$m, as this spectral domain is already covered by spectral order 1 \citep{albert_near_2023} and removing these redder wavelengths also prevents contamination from order 1. Given the lower SNR and the fewer native-resolution channels in order 2, we only performed a broadband light curve fit for this order. We then normalized each light curve by its median and performed 10 iterations of $3.5$-$\sigma$ clipping to remove outliers in flux.

\subsection{Broadband and Spectroscopic Light Curve Fitting} \label{subapp:lcfit}

We first fitted the broadband light curves two visits independently with ExoTEP \citep{benneke_spitzer_2017,benneke_sub-neptune_2019,benneke_water_2019} (Figures~\ref{fig:wlc_fit}(a)--\ref{fig:wlc_fit}(c) and \ref{fig:wlc_fit}(i)--\ref{fig:wlc_fit}(k)), which fits a \texttt{batman} transit model \citep{kreidberg_batman_2015} and a systematics model to the light curve, along with a photometric noise parameter, using the \texttt{emcee} implementation \citep{foreman-mackey_emcee_2013} of the Markov Chain Monte Carlo (MCMC) method, with priors listed in Table~\ref{tab:priors_posteriors_lcfit}. We applied a Gaussian prior on the impact parameter based on the value and uncertainty from \citet{agol_refining_2021} given the low temporal resolution during the transit ingress and egress. For the GP lengthscale (Appendix~\ref{subapp:systematics}), we applied a lower bound corresponding to about 5 integrations to prevent overfitting, and an upper bound of 2.4\,hours, approximately half the duration of the observation, to prevent degeneracies with the parameters of the linear function (Appendix~\ref{subapp:systematics}). These independent fits of the two broadband light curves provided a first estimate of the global transit parameters ($a/R_\star$, $b$), transit ephemeris ($T_{\rm mid}$), and transit depths in the SOSS bandpass. We then fitted the two broadband light curves simultaneously (posteriors listed in Table~\ref{tab:priors_posteriors_lcfit}), allowing each visit to have different systematics and photometric noise parameters, initializing all parameters at their best-fit values from the individual fit to ease convergence. 

\begin{deluxetable*}{lcc}
        \tabletypesize{\scriptsize}
        \tablewidth{0pt} 
        \tablecaption{Priors and posteriors of the joint broadband light curve fit of the two visits.}
        \tablehead{
        \multicolumn{3}{c}{White Light Curve} \\
        \colhead{Parameter} & \colhead{Priors} & \colhead{Posteriors}
        } 
        \startdata 
        Orbital period, $P_{\rm orb}$ (days) & $1.51087081$ & $1.51087081$ \\
        Mid-transit time, visit 1, $T_{\rm mid, 1}$ (BJD) & $\mathcal{U}(2459779.1, 2459779.3)$ & $2459779.210475_{-0.000025}^{+0.000025}$ \\
        Mid-transit time, visit 2, $T_{\rm mid, 2}$ (BJD) & $T_{\rm mid, 1} + 1.51087081$ & $T_{\rm mid, 1} + 1.51087081$\\
        Planet-to-star radius ratio, $(R_p/R_s)$ & ... & $0.08501_{-0.00049}^{+0.00047}$ \\
        Planet-to-star radius ratio squared, $(R_p/R_s)^2$ (ppm) & $\mathcal{U}(100, 50000)$ & $7226_{-83}^{+80}$ \\
        Impact parameter, $b$ & $\mathcal{N}(0.095, 0.65)$ & $0.098_{-0.045}^{+0.046}$ \\
        Quadratic LD coeff. 1, $q_1$ & $\mathcal{U}(0, 1)$ & $0.409_{-0.081}^{+0.101}$ \\
        Quadratic LD coeff. 2, $q_2$ & $\mathcal{U}(0, 1)$ & $0.266_{-0.072}^{+0.080}$ \\
        Eccentricity, $e$ & $0$ & $0$ \\
        Argument of periapsis, $\omega$ (deg) & $90$ & $90$ \\
        Semi-major axis, $a/R_\star$ & $\mathcal{U}(10, 30)$ & $20.66_{-0.15}^{+0.12}$ \\
        White noise, visit 1, $s_1$ (ppm) & $\mathcal{U}(10^{-10}, 10^{10})$ & $172.3_{-11.2}^{+12.3}$ \\
        White noise, visit 2, $s_2$ (ppm) & $\mathcal{U}(10^{-10}, 10^{10})$ & $164.10_{-9.89}^{+10.86}$ \\
        GP amplitude, visit 1, $\log_{10}(a_{\rm GP, 1})$ & $\mathcal{U}(-10, 1)$ & $-3.79_{-0.18}^{+0.26}$ \\
        GP amplitude, visit 2, $\log_{10}(a_{\rm GP, 2})$ & $\mathcal{U}(-10, 1)$ & $-3.68_{-0.11}^{+0.14}$ \\
        GP lengthscale, visit 1, $\log_{10}(\lambda_{\rm GP, 1} ({\rm days}))$ & $\mathcal{U}(-2.21, -1)$ & $-1.62_{-0.31}^{+0.18}$ \\
        GP lengthscale, visit 2, $\log_{10}(\lambda_{\rm GP, 2} ({\rm days}))$ & $\mathcal{U}(-2.21, -1)$ & $-2.01_{-0.12}^{+0.28}$ \\
        \\
        \hline
        \multicolumn{3}{c}{Spectroscopic Light Curves} \\
        \colhead{Parameters} & \multicolumn{2}{c}{Priors} \\
        \hline
        Orbital period, $P_{\rm orb}$ (days) & \multicolumn{2}{c}{$1.51087081$} \\
        Mid-transit time, $T_{\rm mid}$ (BJD) & \multicolumn{2}{c}{WLC best-fit} \\
        Planet-to-star radius ratio squared, $(R_p/R_s)^2$ (ppm) & \multicolumn{2}{c}{$\mathcal{U}(100, 50000)$} \\
        Impact parameter, $b$ & \multicolumn{2}{c}{WLC best-fit} \\
        Quadratic LD coeff. 1, $q_1$ & \multicolumn{2}{c}{$\mathcal{U}(0, 1)$} \\
        Quadratic LD coeff. 2, $q_2$ & \multicolumn{2}{c}{$\mathcal{U}(0, 1)$} \\
        Eccentricity, $e$ & \multicolumn{2}{c}{$0$} \\
        Argument of periapsis, $\omega$ (deg) & \multicolumn{2}{c}{$90$} \\
        Semi-major axis, $a/R_\star$ & \multicolumn{2}{c}{WLC best-fit} \\
        White noise (ppm) & \multicolumn{2}{c}{$\mathcal{U}(10^{-10}, 10^{10})$} \\
        GP amplitude, $\log_{10}(a_{\rm GP})$ & \multicolumn{2}{c}{$\mathcal{U}(-10, 1)$} \\
        GP lengthscale, $\log_{10}(\lambda_{\rm GP} ({\rm days}))$ & \multicolumn{2}{c}{$\mathcal{U}(-2.21, -1)$} \\
        \enddata
        \tablecomments{For the posteriors, the leftmost value is the $50^{\rm th}$ percentile of the posterior distribution, and upper and lower uncertainties are computed from the $16^{\rm th}$, $50^{\rm th}$, and $84^{\rm th}$ percentiles. Posteriors are not listed for the spectroscopic light curves as they are different for each spectral bin.
        }
        \label{tab:priors_posteriors_lcfit}
\end{deluxetable*}

\begin{figure*}
    \centering
    \includegraphics[width=\textwidth]{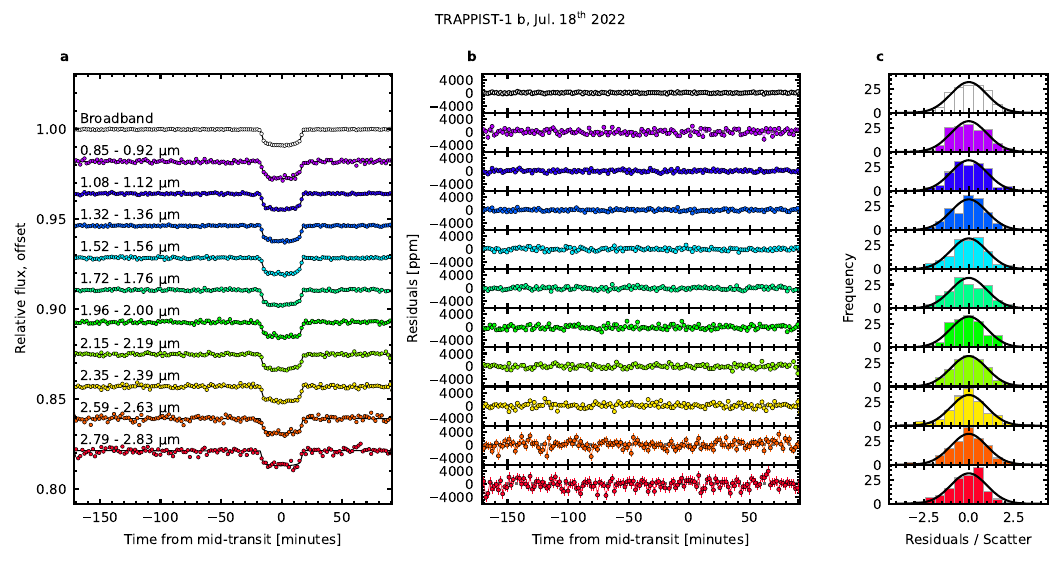}
    \includegraphics[width=\textwidth]{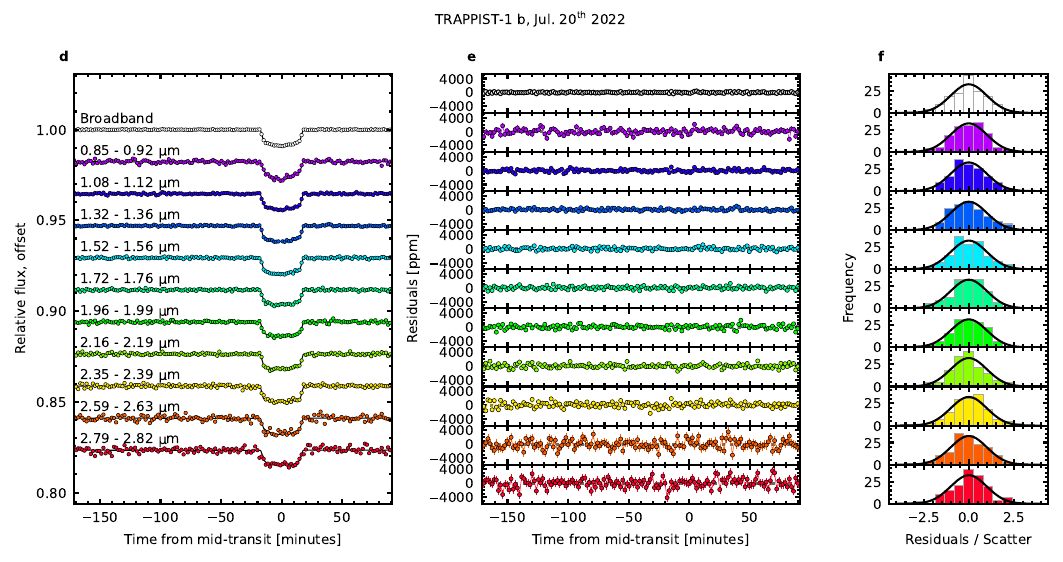}
    \caption{NIRISS/SOSS broadband (white) and spectrophotometric (colored) transit light curves. Panels (a)--(c) correspond to visit 1 and (d)--(f), to visit 2. All error bars are the 1-$\sigma$ uncertainties. (a) \& (d) Light curves. (b) \& (e) Light curve residuals from the best-fit model. (c) \& (f) Distribution of the light curve residuals, with a Gaussian distribution overplotted for comparison.}
    \label{fig:lightcurves}
\end{figure*}

We fitted the spectroscopic light curves of the two transits independently, with a transit and systematics model as well as a photometric noise parameter for each wavelength bin (Figure~\ref{fig:lightcurves}), with priors listed in Table~\ref{tab:priors_posteriors_lcfit}. We parameterized the quadratic limb darkening parameters following \citet{kipping_efficient_2013}. The transmission spectrum for each visit (Figures~\ref{fig:tspec}(a)--\ref{fig:tspec}(b)) is given by the best-fit spectroscopic $R_{\rm p}/R_\star$ values squared.

\subsection{Systematics Treatment} \label{subapp:systematics}

The broadband light curve of the second visit (Figure~\ref{fig:wlc_fit}(i)) betrays the presence of 3- to 15-minute-long correlated noise structures produced by stellar activity (Section~\ref{sec:lcfit}). This correlated noise is most easily seen as flux variations before and after the transit egress of the second visit, but more subtle modulations are also present elsewhere in both visits. To mitigate the impact of this correlated noise on the measured transit depths, we tested two different systematics treatments for both transits, and three additional treatments for the second visit, specifically to leverage the apparent correlation between the H$\alpha$ integrated flux and the flux in other wavelength bins. We tested each of these five systematics treatments in parallel during the transit light curve fitting (Appendix~\ref{subapp:lcfit}). In all five systematics treatments, we used a linear function as the basic systematics model $S(t)$, such that the most basic flux model is written as
\begin{equation}
\begin{aligned}
    f(t) = f_0 \cdot T(t, \theta) \cdot S(t), \, {\rm with} \ S(t) = v \cdot (t - t_0) + c\,,
\end{aligned}
\end{equation}
where $f_0$ is the baseline flux, $T(t, \theta)$ is the transit model, $v$ and $c$ are the slope and normalization constant of the linear function, and $t_0$ is the time at the start of the visit.

The first systematics treatment consisted in trimming the baseline and masking temporally correlated points such that the most obviously correlated noise patterns would not affect the fit. For visit 1, we removed 2 and 0.6 hours from the start and the end of the visit, respectively. For visit 2, we masked 6 in-transit and 12 post-transit points, all of which seemed correlated in time and with H$\alpha$. We also removed 2 hours from the start, but did not remove any additional parts from the end, since 12 points were already masked, such that trimming would have left very little post-transit baseline. 

The second systematics treatment consisted in using a Gaussian process (GP) \citep{rasmussen_gaussian_2006}. We used a squared-exponential kernel to define the covariance matrix of the likelihood function using the \texttt{george} Python package \citep{ambikasaran_fast_2015}, keeping the white photometric noise parameter on the diagonal of the covariance matrix, added in quadrature to the kernel function. This added two free hyperparameters to the fit of each light curve, the amplitude $a_{\rm GP}$ and timescale $\lambda_{\rm GP}$ of the GP. 

The subsequent systematics treatments were only applied to the second visit as they all involved the H$\alpha$ luminosity, which was most obviously correlated with the flux in other wavelengths for this visit. The third systematics treatment consisted in detrending the light curves against the H$\alpha$ integrated flux, that is, 
\begin{equation}
    S(t) = v \cdot (t - t_0) + c + w \cdot (F_{\rm H\alpha}(t) - F_{\rm H\alpha, 0})\,,
\end{equation}
where $w$ is a free parameter for each light curve, $F_{\rm H\alpha}(t)$ is the H$\alpha$ luminosity time series and $F_{\rm H\alpha, 0}$ is the H$\alpha$ luminosity at the start of the visit.

In the fourth systematics treatment, we first trained a squared-exponential kernel GP on the H$\alpha$ luminosity time series, 
yielding a Gaussian-like posterior distribution function (PDF) for both $a_{\rm GP}$ and $\lambda_{\rm GP}$. We used the mean and standard deviation of the PDF of $\lambda_{\rm GP}$ to define a Gaussian prior applied on $\lambda_{\rm GP}$ as we fitted the broadband and spectroscopic light curves with a squared-exponential kernel GP.

The fifth systematics treatment was a combination of the second and third treatments, that is, detrending against the H$\alpha$ luminosity \textit{and} using a (non-trained) squared-exponential kernel GP.

Transit depths are consistent within their uncertainties across all systematics treatments (Figure~\ref{fig:tspec_pipelines_systematics}(b)). This shows that the GP does not bias the measured $R_{\rm p}/R_{\rm s}$, at least not in a way that is different from simply masking parts of the light curves or simply detrending against the H$\alpha$ luminosity. The median error bars, in parentheses in the legend, are larger for the trimmed baseline treatment and any treatment involving a GP. The larger error bars for the trimmed baseline treatment is simply explained by the 
reduced amount of data, especially in visit 2 where in-transit points were masked. We explain the larger error bars in the GP treatments by the fact that the GP hyperparameters may be correlated with the transit depths, which inflates their error bars. However, a smaller median error bar on the transit depths does not necessarily mean a better fit to the light curves. A visual inspection of the light curve residuals from the best-fit model of each wavelength bin for each systematics treatment indicated that the uninformed GP (treatment 2) provided a better fit, that is, it yielded residuals with the least correlated noise. This is expected given its more flexible nature compared to detrending against the fixed H$\alpha$ time series. This is why we selected treatment 2 as our reference.

\section{Stellar variability} \label{app:stelvar}

The extracted light curves of the two visits showed numerous indicators of stellar variability. The average out-of-transit linear slope of the broadband light curve of visit 1 was $8.21\pm0.24\,$ppm\,min$^{-1}$, within the range of values inferred from the \textit{K2} light curve \citep[$\pm10\,$ppm\,min$^{-1}$, computed from][]{luger_seven-planet_2017}. For visit 2, the out-of-transit baseline had a curvature that could not be properly described by a linear function, consistent with an inflection point in the $K2$ light curve. Since the two visits were approximately $1.5\,$days apart \citep{gillon_seven_2017}, and assuming the stellar flux is modulated with a period of $3.3\,$days \citep{luger_seven-planet_2017}, it would be possible for the first visit to be in a phase of increasing stellar flux and the second, at an inflection point. During commissioning, two other stars showed different slope values in their light curves \citep{doyon_near_2023}, suggesting that the flux variations seen in the TRAPPIST-1 data are \textit{astrophysical}, not instrumental. 

\begin{figure*}
    \centering
    \includegraphics[width=1.0\textwidth]{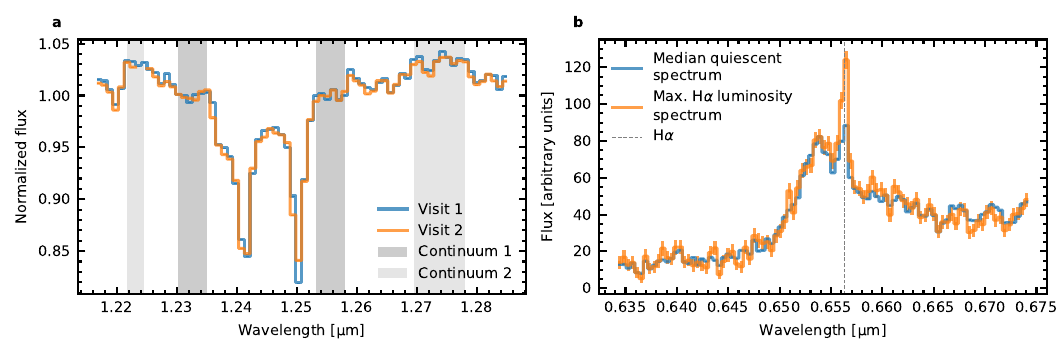}
    \caption{Signatures of stellar variability and activity between and within the two visits. (a) Median extracted spectrum of visits 1 (blue) and 2 (orange), zoomed in on the 1.24--1.25\,$\mu$m K\textsc{i} doublet. Uncertainties are smaller than the line width. Shaded regions show the two different definitions of the continuum adopted to compute the equivalent width of the reddest of the two lines of the doublet. (b) Median quiescent (out-of-flare) stellar spectrum of visit 2 (blue) and stellar spectrum at maximum H$\alpha$ luminosity (orange), i.e. at the peak of the flare in visit 2. Error bars are the 1-$\sigma$ uncertainties.}
    \label{fig:stellaractivity}
\end{figure*}

The temperature-sensitive K\textsc{i} doublet at $1.24$--$1.25\,\mu$m \citep{geballe_infrared_2001,fuhrmeister_carmenes_2022} was detected in both visits (Figure~\ref{fig:stellaractivity}(a)). The average equivalent width (EW) of the reddest of the two lines is $4.134\pm0.003$ and $4.034\pm0.003\,$\AA\, for visits 1 and 2, respectively, adopting the dark shaded region in Figure~\ref{fig:stellaractivity}(a) as the continuum. Using the light shaded region in Figure~\ref{fig:stellaractivity}(a) as the continuum instead, we measured $5.510\pm0.003$ and $5.338\pm0.003\,$\AA. We used PHOENIX ACES models \citep{husser_new_2013} to convert this temporal variation of K\textsc{i} EW into a variation in effective temperature to infer that the average photosphere of visit 1 was, depending on our definition of the continuum, $15.3\pm0.6\,$ or $19.7\pm0.4\,$K cooler than visit 2. We adopt $18\pm3\,$K as the average photosphere temperature variation between visits 1 and 2, roughly consistent with the photometric peak-to-peak variability of about $2\,\%$ of the \textit{K2} light curve \citep{luger_seven-planet_2017}. 

The H$\alpha$ stellar line was also detected and showed significant variability within and between visits (Figures~\ref{fig:wlc_fit}(d), \ref{fig:wlc_fit}(l), \ref{fig:stellaractivity}(b)). Continuum flux was correlated with H$\alpha$ as evidenced by the small flare event in the broadband light curve of visit 2, near the egress (compare Figures~\ref{fig:wlc_fit}(j) and \ref{fig:wlc_fit}(l)). Such continuum variation is a clear source of systematic uncertainty for transit depth measurements, hence the five systematics treatments tested (Appendix~\ref{subapp:systematics}). More TRAPPIST-1 observations may be used to derive an empirical calibration between H$\alpha$ and the continuum flux to correct light curves from stellar activity signals, which will likely be very common in this system. 

\section{Stellar spectrum} \label{app:stelspec}

\begin{figure*}
    \centering
    \includegraphics[width=.8\textwidth]{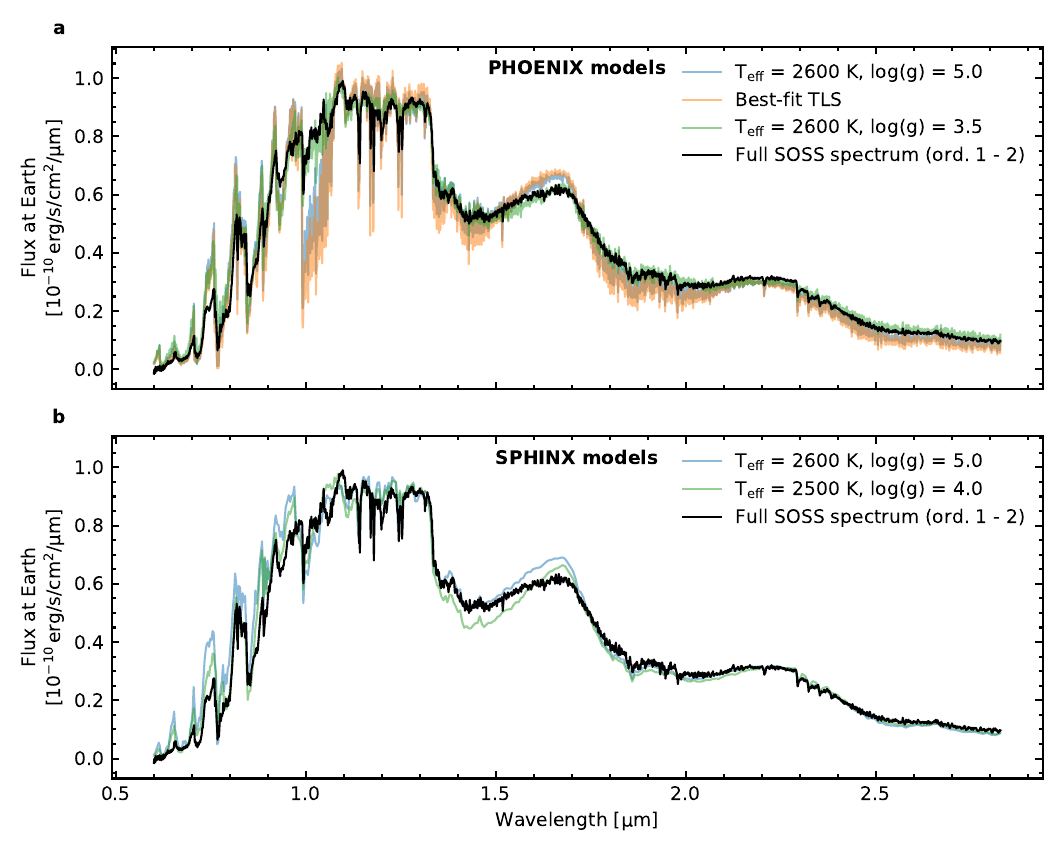}
    \caption{Comparison between the median out-of-transit SOSS spectrum of visit 2 (black) to PHOENIX ACES models \citep{husser_new_2013} (colored curves in panel (a)), and SPHINX models \citep{iyer_sphinx_2023,iyer_sphinx_2023_7416042} (colored curves in panel (b)), normalized to match the SOSS observations in the J band. The blue curves are the models at the effective temperature and surface gravity of TRAPPIST-1 adopted from the literature \citep{agol_refining_2021} (values rounded to $T_{\rm eff}=2600$\,K and $\log{g}=5.0$). In panel (a), the orange curve is a linear combination of three PHOENIX models, where the weight, surface gravity, and temperature of each model is taken from the stellar contamination model fit to the transit spectrum of visit 2 (Section~\ref{subsec:sequential}). The green curves are the models that yielded the best fit between the observed SOSS spectrum and the PHOENIX or SPHINX model spectrum ($T_{\rm eff}=2600$\,K, $\log{(g)}=3.5$ for PHOENIX, $T_{\rm eff}=2500$\,K, $\log{(g)}=4.0$ for SPHINX). For PHOENIX models, we scanned effective temperatures between $2300$ and $2800$\,K with $100$-K increments, and $\log{(g)}$ values between $2.5$ and $5.5$ with $0.5$-dex increments. For SPHINX models, we scanned effective temperatures between $2400$ and $2700$\,K with $100$-K increments, and $\log{(g)}$ values between $4.0$ and $5.5$ with $0.5$-dex increments.}
    \label{fig:stelspectrum_vs_model}
\end{figure*}

To expand on the discrepancies between theoretical stellar spectra and observations, we compared a selection of PHOENIX ACES models \citep{husser_new_2013} to the median, out-of-transit, flux-calibrated SOSS spectrum from visit 2 (Figure~\ref{fig:stelspectrum_vs_model}(a)). To flux-calibrate the spectrum of TRAPPIST-1, we extracted the TRAPPIST-1 spectra, then ran the \texttt{photomStep} of the \texttt{jwst} pipeline using a reference file provided by Kevin Volk (priv. comm.). An extraction box size of 40 pixels was used to match the aperture size used for the A-type star used to build the reference file. 

First, we adopted literature values of the effective temperature and surface gravity of TRAPPIST-1, rounded to the nearest hundred kelvin and $0.5\,$dex, respectively, that is, $T_{\rm eff} = 2566\,K \sim 2600\,K$, $\log{(g)} = 5.2396 \sim 5.0\,$dex \citep{agol_refining_2021}. The PHOENIX model at these literature values, normalized to match the SOSS spectrum in the J band, shown in blue in Figure~\ref{fig:stelspectrum_vs_model}(a), is in disagreement with the observation between $1.0$ and $1.1\,\mu$m, and the general shape of the model is also inconsistent with the data. 

Second, we used the best-fitting spot and facula covering fractions, temperatures, and surface gravities from the stellar contamination fit on the transit spectrum of the second visit (from the sequential fit, Table~\ref{tab:priors_posteriors_tls}) to compute a linear combination of PHOENIX models as described in Section~\ref{subsec:sequential}. This TLS-informed spectrum, also normalized to match the SOSS data in the J band and shown in orange in Figure~\ref{fig:stelspectrum_vs_model}(a), exhibits differences with the observed spectrum similar to the model at literature values. 

Third, we fitted the spectrum with linear combinations of PHOENIX spectra following \citet{rackham_towards_2023}, testing models with 1, 2, 3, and 4 photospheric components. These best-fit models are not shown in Figure~\ref{fig:stelspectrum_vs_model} for clarity, but their fit to the observed spectrum is similar to that of the literature spectrum and that of the best-fit TLS spectrum, i.e., poor. The Bayesian evidence suggests preference for the three-component model, though the data residuals are still considerable and the flux-calibrated data uncertainties need to be inflated to 22\% of the observed flux to achieve a reduced chi-squared ${\sim}1$. Assuming the planet transits the dominant spectral component and calculating a TLS correction while propagating the uncertainty of our posterior parameter estimates, we find that this limitation in model fidelity imparts a model-driven uncertainty on the corrected transmission spectrum of ${\sim}1800$\,ppm per native-resolution pixel. In line with previous analyses of TRAPPIST-1 transmission spectra \citep{zhang_near-infrared_2018, wakeford_disentangling_2019, garcia_hstwfc3_2022}, this result suggests that our understanding of TRAPPIST-1's heterogeneous photosphere and limitations with current approaches for modeling such a photosphere drive the ultimate precision of our transmission measurements in this system. 

Last, we scanned through a grid of PHOENIX models with temperatures ranging from $2300$ to $2800\,$K in steps of $100\,$K and surface gravities from $2.5$ to $5.5\,$dex in steps of $0.5\,$dex. For each $(T, \log{g})$ combination, we binned the PHOENIX model down to the NIRISS wavelength grid, normalized the model to match it to the NIRISS spectrum in the J band, and computed the $\chi^2$ between the PHOENIX model and the NIRISS spectrum. The model yielding the smallest $\chi^2$ was at $T=2600\,$K and $\log{(g)}=3.5\,$dex, shown in green in Figure~\ref{fig:stelspectrum_vs_model}(a). This model provides a better fit to the observation than the literature and the TLS-informed models, but still cannot fit certain features, e.g., from $0.7$ to $0.8\,\mu$m. 
The best-fitting surface gravity is very low and we suspect this is only because it mimics the effect of other physical processes (e.g. magnetic fields) not captured in the simple $T$-$\log{g}$ parameter space; it is unlikely that the entire star or even large surfaces of it would actually have such a low surface gravity. Other signs of low gravity were previously reported \citep{burgasser_age_2017} based on low-resolution near-infrared spectra \citep{gillon_temperate_2016}. It was suggested that the presence of magnetic fields and/or tidal interactions with the planets in the system could explain these low-gravity spectral features \citep{gonzales_reanalysis_2019}. 

We also compared the SOSS spectrum to SPHINX models \citep{iyer_sphinx_2023} in a similar fashion, with the analogous models presented in Figure~\ref{fig:stelspectrum_vs_model}(b). We did not compute a linear combination of SPHINX models constructed from the best-fit TLS parameters, as the TLS fit was performed with PHOENIX models. The SPHINX models generally provide a better fit to the SOSS spectrum in the $0.9$--$1.1\,\mu$m domain compared to the PHOENIX models, but at the expense of the $1.5$--$1.7\,\mu$m domain, where the SPHINX models struggle to reproduce the observed water band wing. Given the lower spectral resolution of the SPHINX models ($R\sim250$) and their limited $\log{g}$ values available ($4.0$ to $5.5\,$dex), the TLS fits were performed with PHOENIX models only.

In summary, none of the models we tested can reproduce the observed TRAPPIST-1 spectrum over the entire SOSS bandpass at the same level they would an earlier-type star. Because model fidelity is required to correct for stellar contamination in transmission spectra \citep{rackham_towards_2023}, the mismatch between state-of-the-art models and observations call for theoretical developments to cover the missing physics and/or more observations of the star to correct for contamination in a data-driven approach.

\section{Simultaneous Stellar Contamination and Planetary Atmosphere Fits Supplemental Information} \label{app:corner}

\begin{figure*}
    \centering
    \includegraphics[width=\textwidth]{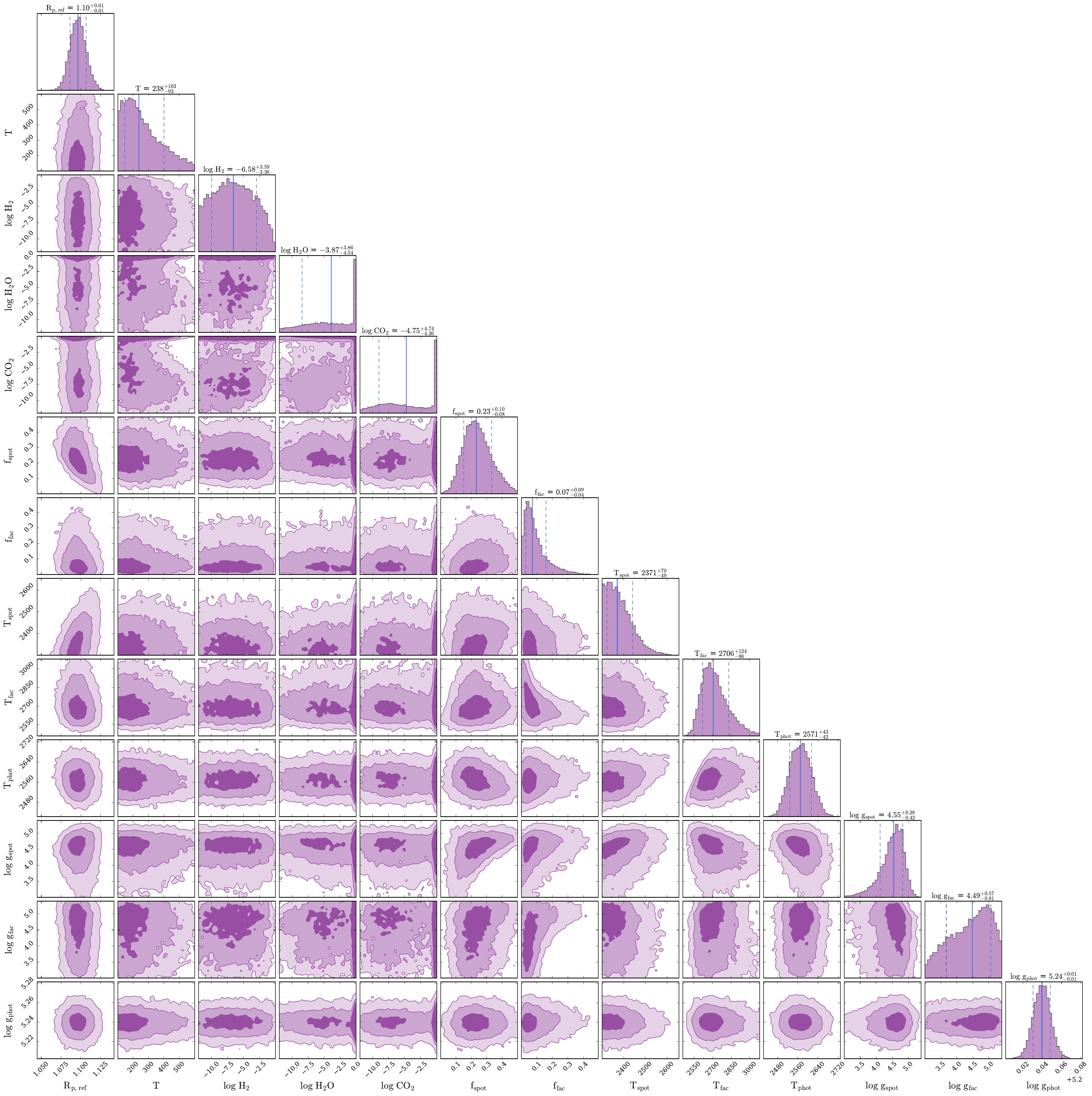}
    \caption{\texttt{POSEIDON} posterior distribution for visit 1. This corresponds to the simultaneous fit of the stellar contamination and planetary atmosphere for the first TRAPPIST-1~b transit (see Section~\ref{subsec:simultaneous}).}
    \label{fig:corner_poseidon_vis1}
\end{figure*}

\begin{figure*}
    \centering
    \includegraphics[width=\textwidth]{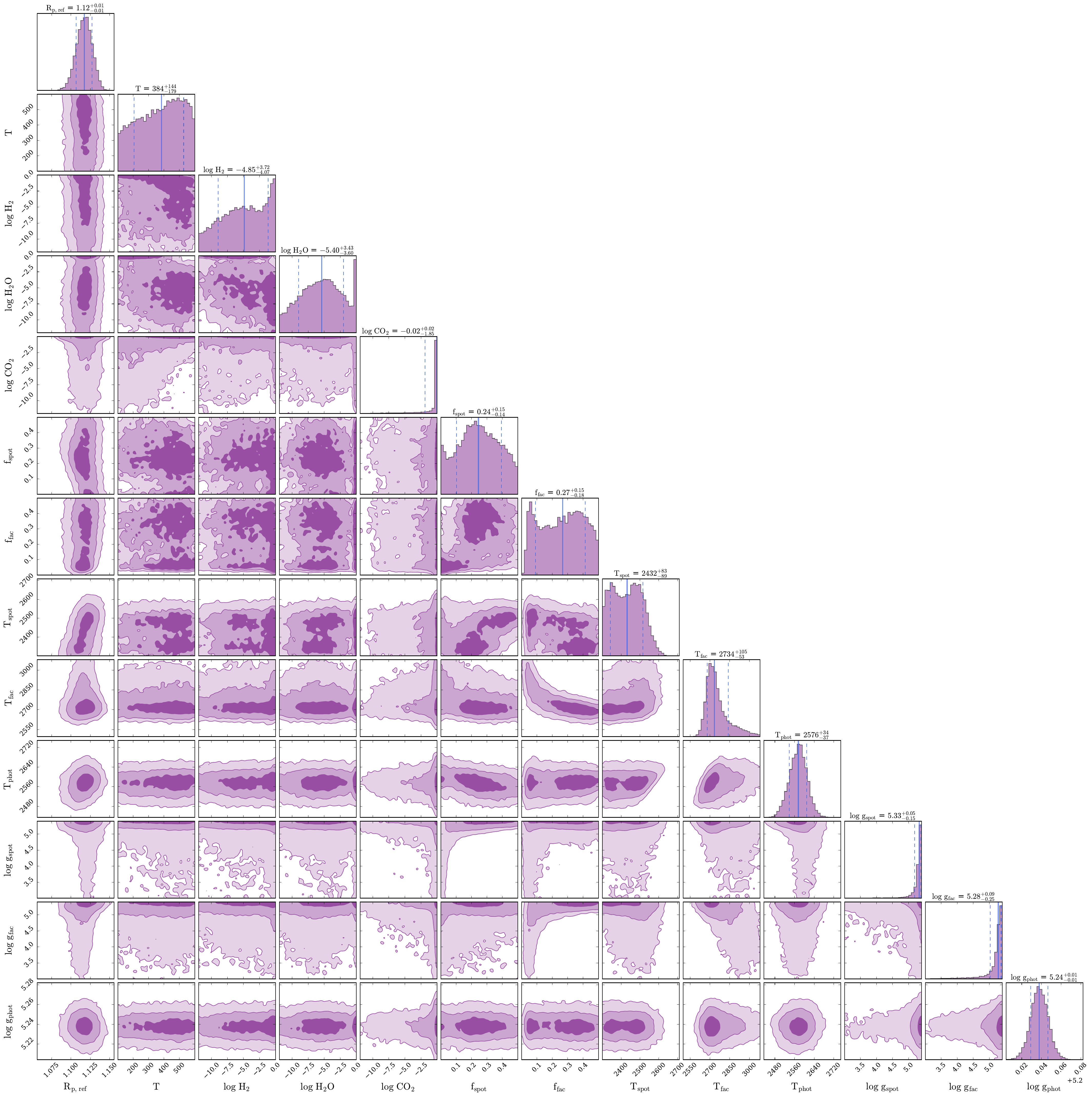}
    \caption{\texttt{POSEIDON} posterior distribution for visit 2. This corresponds to the simultaneous fit of the stellar contamination and planetary atmosphere for the second TRAPPIST-1~b transit (see Section~\ref{subsec:simultaneous}).}
    \label{fig:corner_poseidon_vis2}
\end{figure*}

This appendix provides the full corner plots resulting from the simultaneous fits of the stellar contamination and planetary atmosphere with \texttt{POSEIDON} for visits 1 and 2 (Figures~\ref{fig:corner_poseidon_vis1} and \ref{fig:corner_poseidon_vis2}).

\bibliography{zotero_ol,extra}{}
\bibliographystyle{aasjournal}



\end{document}